 \documentclass[12pt]{article}
 \usepackage{amsfonts,amssymb}
 \newcommand{\End}{\nonumber\\ }
 \newcommand{\Proof}{{\bf Proof}\ \ }
 \newcommand{\Oproof}{{\bf Outline of proof}\ \ }
 \newcommand{\Endproof}{\hfil $\blacksquare$ \par}
 \newcommand\Mbox[1]{ \mbox{{\rm #1}}}
 \newcommand\Real{ {\mathbb R} }
 \newcommand\Comp{ {\mathbb C} }
 \newcommand{\Frac}[2]{{\textstyle{\frac{#1}{#2}}}}
 \newcommand\Half{\Frac{1}{2}}
 \newcommand{\Sum}[2]{\sum_{#1}^{#2}}
 \newcommand{\Defi}[1]{{\em #1}}
 \newcommand{\Gp}[1]{\epsilon_{#1}}
 \newcommand{\Cfac}[2]{(-1)^{(\Gp{\Sscap{#1}} \Gp{\Sscap{#2}}) }}
 \newcommand{\Rs}{\Real_S}
 
 \newcommand{\Rstt}[2]{\Real_{S}^{#1,#2}}
 \newcommand{\Srow}[4]{(#1^1,\dots,#1^{#2};#3^1,\dots,#3^{#4})}
 
 \newcommand{\Bint}{{}_{_{\mathcal B}}\!\!\!\!\int}
  \newcommand{\Sint}[1]{{}_{_{\mathcal S}}\!\!\!\!\int_{#1}}
 \newcommand{\Num}[1]{|#1|}
 \newcommand{\Df}{{\rm d}}
 \newcommand{\Fclass}{\mathcal{F}}
 \newcommand{\Exc}{\mathbb{E}_c}
 \newcommand{\Excb}[1]{\mathbb{E}_c{ \left[ #1 \right]}}
 \newcommand{\Exg}[1]{\mathbb{E}_G{ \left[ #1 \right]}}
 \newcommand{\Exgb}[1]{\mathbb{E}_{\mathcal{G}}{ \left[ #1 \right]}}
 \newcommand{\Exs}{\mathbb{E}_{\mathcal{S}}}
 \newcommand{\Exsb}[1]{\mathbb{E}_{\mathcal{S}}{ \left[ #1 \right]}}
 \newcommand{\Muq}{=_{\mathbb{E}}{}}
 \newcommand{\Ker}[1]{[{#1}]}
 \newcommand{\Ft}[1]{\hat{#1}}
 \newcommand{\Exp}[1]{\exp \left( #1 \right)}
 \newcommand{\Expi}[1]{\exp i \left( #1 \right)}
 \newcommand{\Dbd}[1]{\Frac{\partial}{\partial #1}}
  \newcommand{\Dbdf}[2]{\frac{\partial #1}{\partial #2}}
  \newcommand{\Ds}[1]{\partial_{#1}}
 \newcommand{\One}{\mathbf{1}}
 \newcommand{\Intzt}{\int_0^{t}}
 
 \newcommand{\Chri}[3]{\Gamma_{#1 #2}{}^{#3}}
 \newcommand{\Conn}[3]{A_{#1\, #2}{}^{#3}}
 \newcommand{\Scap}[1]{{\scriptstyle {#1}}}
  \newcommand{\Sscap}[1]{{\scriptscriptstyle {#1}}}
 
 \newcommand{\Fsc}{{\Scap{F}}}
 
 \newcommand{\Ind}{\Mbox{Index}}
 \newcommand{\Tr}{\Mbox{Trace}}
 \newcommand{\Str}{\Mbox{Supertrace}}
 \newcommand{\tr}{\Mbox{tr}}
 \newcommand{\str}{\Mbox{str}}
 \newcommand{\Matt}[4]{\left(
 \begin{array}{cc}#1&#2\\#3&#4\end{array}\right)}
  \newcommand{\Hf}{L_0}
 \newcommand{\Hs}{L_1}
 \newcommand{\Ho}{L_2}
 
 \newcommand{\Mproj}[1]{\left\{ #1 \right\}_{m}}
 \newcommand{\Brst}{\Omega}
 \newcommand{\Gff}{\chi}
 \newcommand{\Com}[2]{{}[ #1 , #2 ]}
 \newcommand{\Hamg}{H_g}
 \newcommand{\Hame}{H_{ext}}
 \newcommand{\Comm}{\Com{\Brst}{\Gff}}
 \newcommand{\Lag}{\mathcal{L}}
 \newcommand{\Pb}[2]{\left\{ #1 , #2 \right\}}
 \newcommand{\Xt}{\tilde{x}}
 \newcommand{\Cj}[1]{{}* #1}
 \newcommand{\Msbox}[1]{\Mbox{\ {\tiny{#1}}\ }}
 \newcommand{\Sm}[2]{{\mathcal{S}}(#1,#2)}
 \newcommand{\Som}{\Sm{O(M)}{T(M)\otimes E}}
 \newcommand{\Somu}{\Sm{O(M)}{T(M)}}
 \newcommand{\Wt}{\widetilde{W}}
 \newcommand{\Wi}{\gamma}
 
 \newcommand{\Spinsm}{\Sm{M}{E_{O(M)}}}
 \newcommand{\Dira}{\not\!\! D}
 \newcommand{\Mm}[3]{M^{#1}_{#2 \, #3}}
 \newcommand{\Subtt}{{}_{(t;\tau)}}
 \newcommand{\Subss}{{}_{(s;\sigma)}}
 \newcommand{\Intzztt}{\int_0^{\tau} \int_0^t}
 \newcommand{\Dss}{D[s;\sigma]}
 \newcommand{\Psib}{\overline{\psi}}
 \newcommand{\Ltr}[1]{\mathcal{L}^2(\Real^{#1})}
 \newcommand{\Ga}{{{}_{[a]}}}
 \newcommand{\Gb}{{{}_{[b]}}}
 \newcommand{\Gab}{{}_{{}_{\Gamma_{ab}}}}
 \newcommand{\Hamu}{H_u}
 \newcommand{\Dctwoh}[3]{D_{#2}D_{ #3}h(#1)}
 \newcommand{\xm}{\tilde{x}}
 \newcommand{\Intdt}{\int_0^{\delta t}}
 \newcommand{\Meh}{M}
 \newcommand{\Fmpos}{\tilde{\Fpos}}
 \newcommand{\Fmmom}{\tilde{\Fmom}}
 \newcommand{\Fpos}{\theta}
 \newcommand{\Fmom}{\rho}
 \newcommand{\Curv}[4]{R_{#1 #2 #3}{}^{#4}}
 \newcommand{\Csh}{\frac{\cosh us}{\sinh  us}}
 \newcommand{\Dut}{d_{u\,(2)}}
 \newcommand{\psit}{\tilde{\psi}}
 \newcommand{\Hstar}[1]{{}\,\,{}^* #1}

 \newcommand{\Dp}[2]{\frac{\partial #1}{\partial #2}}
\newtheorem{Def}{Definition}[section]

\newtheorem{The}[Def]{Theorem}
\newtheorem{Lem}[Def]{Lemma}
\newtheorem{Exa}[Def]{Example}

\begin{document}
 \begin{center}
 {\large Supersymmetry and Brownian motion on supermanifolds}\\
 \ \\
   Alice Rogers\\
  \ \\
  Department of Mathematics      \\
  King's College             \\
  Strand, London  WC2R 2LS         \\
 \end{center}

 \vskip0.2in
 \begin{center}
 December 2001
 \end{center}
 \vskip0.2in
 \begin{abstract}
An anticommuting analogue of Brownian motion, corresponding to
fermionic quantum mechanics, is developed, and  combined with
classical Brownian motion to give a generalised  Feynman-Kac-It\^o
formula for paths in geometric supermanifolds.  This formula is
applied to give a rigorous version of the proofs  of the
Atiyah-Singer index theorem based on supersymmetric quantum
mechanics. It is also shown how superpaths, parametrised by a
commuting and an anticommuting time variable, lead to a manifestly
supersymmetric approach to the index of the Dirac operator. After
a discussion of the BFV approach to the quantization of theories
with symmetry, it is shown how the quantization of the topological
particle leads to the supersymmetric model introduced by Witten in
his study of Morse theory.
 \end{abstract}
 \section{Introduction}
This survey concerns a battery of generalised probabilistic
techniques, originally motivated by path integration in fermionic
and supersymmetric quantum physics, which may be brought to bear
on some significant geometric operators.

A theory of Brownian motion in spaces with anticommuting
coordinates is developed, together with the corresponding
stochastic calculus. This is combined with conventional Brownian
motion  to provide a mathematically rigorous version of the direct
and intuitive proofs of the Atiyah-Singer index theorem based on
supersymmetric quantum mechanics \cite{Alvare,FriWin}, and also of
the  quantum tunneling calculations which are necessary to build
Witten's link between supersymmetry and Morse Theory
\cite{Witten82}.

The connection between anticommuting variables and geometry arises
via Clifford algebras.  It seems to have first been observed in
the context of canonical anticommutation relations for fermi
fields (which are essentially Clifford algebra relations) that
such algebras have a simple representation in terms of
differential operators on spaces of functions of anticommuting
variables.  It was thus in the physics literature that functions
of anticommuting variables were first considered, beginning with
the work of \cite{Martin} and ideas of Schwinger \cite{Schwin},
and extensively developed by Berezin \cite{Berezi1966} and by
DeWitt \cite{Dewitt1984}.

Anticommuting variables are not used to model physical quantities
directly; their use is motivated by the algebraic properties of
the function spaces of these variables. In application to physics,
results which are real or complex numbers emerge after what has
become known as  Berezin integration (defined in equation
(\ref{BINTeq})) which essentially takes a trace. The approach to
fermions  using anticommuting variables is particularly useful in
the context of supersymmetry  symmetry because bose and fermi
degrees of freedom, which are related by symmetry transformations,
can then be handled in  the same way. Similarly, when using
{\textsc{BRST}} techniques to consider theories with symmetry, the
use of anticommuting variables to handle ghost degrees of freedom
provides a unifying approach.

At this stage a comment on the prefix `super' is appropriate.
Originally used in the physical term {\em supersymmetry}, to
describe a symmetry which mingles bosonic and fermionic degrees of
freedom, the word has migrated into mathematics to describe a
generalisation or extension of a classical object to an object
which is in some sense $Z_2$-graded, with an even part usually
associated with the classical, commuting object and an odd part
associated with some anticommuting analogue.

A generic superspace is a space with coordinates some of which are
commuting and others anticommuting.  There are two approaches to
superspace, a concrete one which generalises the actual space, and
an abstract one which generalises the algebra of functions on such
a space.  Here generally the concrete approach will be used
because it allows simpler super analogues of the classical
concepts considered;  however it should be born in mind that the
function algebras used in the concrete approach will normally
correspond to the generalised algebras in the more abstract
approach, so that the difference is more apparent than real.

The anticommuting analogue of probability theory developed here
does not mirror all aspects of classical probability theory, but
is largely restricted to features which relate directly to the
heat kernels of differential operators.

In section \ref{SSSMsec} of this survey superspace is described,
the concept of supersmooth function is defined together with the
related notions of differentiation and integration.   The
construction of a particular class of supermanifold, obtained by
building odd dimensions onto a manifold using the data of a vector
bundle, is also given.  These supermanifolds play a key role in
the geometric applications of the anticommuting probability
theory, because they carry natural classes of supersmooth
functions which correspond to spaces of forms and spinors on the
underlying manifold.  Section \ref{SUSYQMsec} introduces fermionic
and supersymmetric quantum mechanics, explaining quantization in
terms of functions of commuting and anticommuting variables.
Systems are described whose Hamiltonians correspond to a number of
geometric Laplacians.

Section \ref{GPSsec} introduces the anticommuting analogue of
probability measure, and in particular fermionic Wiener measure.
In section \ref{SWMsec} fermionic Wiener measure is combined with
classical Wiener measure to give a super Wiener measure for paths
in superspace. The stochastic calculus for the corresponding super
Brownian paths is developed. Next, in section \ref{BMSMsec}, these
techniques are used to define Brownian paths on supermanifolds and
derive a Feynman-Kac-It\^o formula for certain Hamiltonians.

The concept of supersymmetry implies more than the mere presence
of anticommuting and commuting variables; an odd operator, known
as the supercharge, which squares to the Hamiltonian or Laplacian
of the theory is required.  (In some cases the Hamiltonian is the
sum of squares of more than one supercharge.) Supersymmetric
quantum mechanical models are described in section
\ref{SUSYQMsec}, while geometric examples of supercharges are the
Hodge-De Rham operator and the Dirac operator which are considered
in later sections. Using the `super' stochastic machinery, these
operators can be analysed. In section \ref{ASIsec} the methods are
applied to make rigorous the supersymmetric proofs of the
Atiyah-Singer index theorem given by Alvarez-Gaum\'e \cite{Alvare}
and by Friedan and Windey \cite{FriWin}, while in section
\ref{MTTsec} it is used to verify the key instanton calculation in
Witten's linkage of Morse theory with supersymmetric quantum
mechanics. The Dirac operator is studied in section
\ref{DIRACsec}, where a further super construction, the Brownian
super path parametrised by both a commuting time $t$ and an
anticommuting time $\tau$, is introduced. Anticommuting variables
may also be used to handle the ghost degrees of freedom used in
the quantization of theories with symmetry.  These ideas are
briefly described in section \ref{BRSTsec}, and the used in the
following section to show that quantization of the topological
particle leads to the supersymmetric model used by Witten to
analyse Morse theory.

Other work involving anticommuting analogues of probability measures
includes that of Applebaum and Hudson \cite{AppHud}, Barnet, Streater and
Wilde \cite{BarStrWil}, Haba and Kupsch
\cite{HabKup}.  An alternative approach to considering paths in
superspace is to consider differential forms on loop space, as in the
work of Jones and L\'eandre \cite{JonLea}.

The work in this survey shows how an extended notion of path integral may
be applied to a variety of quantum mechanical systems.  When considering
quantum field theory the standard approach is by functional integrals,
but an interesting alternative approach, involving Brownian motion over
loop groups, and  over torus groups and higher dimensional objects, has
been considered in papers by Brzezniak and Elworthy
\cite{BrzElw}, by Brzezniak and L\'eandre
\cite{BrzLea}, and by L\'eandre \cite{Lea01}.  Here too fermionic analogues
might be possible.
 \section{Superspace and supermanifolds}\label{SSSMsec}
This section introduces the analysis and geometry of anticommuting
variables. As explained in the introduction, we use a concrete approach
to the concept of supermanifold. We begin with some algebraic notions,
leading to the key concept of supercommutative superalgebra.
  \begin{Def}{\rm
A \Defi{super vector space} is a vector space $V$ together with a
direct sum decomposition
 \begin{equation}
   V = V_{0} \oplus V_{1}.
 \end{equation}
The subspaces $V_{0}$ and $V_{1}$ are referred to respectively as
the \Defi{even} and \Defi{odd} parts of $V$.
  }\end{Def}
We will normally consider homogeneous elements, that is elements
$\Scap{X}$ which are either even or odd, with  parity denoted by
$\Gp{\Scap{X}}$ so that $\Gp{\Scap{X}}=i$ if $\Scap{X}$ is in
$\Real_{S,i},i=0,1$. (Arithmetic of parity indices $i=0,1$ is always
modulo $2$.)
 \begin{Def} {\rm
  \begin{enumerate}
  \item A \Defi{superalgebra} is a super vector space $A= A_0 \oplus A_1$
 which is also an algebra which satisfies $A_i A_j \subset A_{i+j}$.
  \item The superalgebra is \Defi{supercommutative} if, for all
   homogeneous $\Scap{X}, \Scap{Y}$ in $A$,
   $\Scap{X}\Scap{Y} = \Cfac{X}{Y} \Scap{Y}\Scap{X}$.
  \end{enumerate}
 }
 \end{Def}
The effect of this definition is that in a superalgebra the
product of an even element with an even element and that of an odd
element with an odd element are both even while the product of an
odd element with an even element is odd;  if the algebra is
supercommutative then odd elements anticommute, and the square of
an odd element is zero.

The basic  supercommutative superalgebra used is the real
Grassmann algebra with generators
 $\One, \beta_1, \beta_2, \dots$ and relations
 \begin{equation}
   \One\beta_i = \beta_i \One= \beta_i, \qquad
   \beta_i \beta_j = -    \beta_j \beta_i.
 \end{equation}
This algebra, which is denoted $\Rs$, is a superalgebra with
 $\Rs := \Real_{S,0} \oplus \Real_{S,1}$
where $\Real_{S,0}$ consists of linear combinations of  products
of even numbers of the anticommuting generators, while
$\Real_{S,1}$ is built similarly from odd products.

It is useful to introduce multi-index notation: for a positive
integer $n$, let $M_n$ denote the set of all multi-indices of the
form
 $\mu := \mu_{1} \ldots \mu_{k}$
with  $1 \leq \mu_{1} < \ldots < \mu_{k} \leq n$ together with the
empty multi-index {\small{$\emptyset$}}; also let $|\mu|$ denote
the length of the multi-index $\mu$, and for any suitable
$n$-component object $\xi^1, \dots, \xi^n$ define
 $\xi^\emptyset := \One$ (the appropriate unit for the objects $\xi^i$) and
 $\xi^\mu := \One \xi^{\mu^1} \dots \xi^{\mu_{|\mu|}}$.
The set $M_{\infty}$ is defined in a similar way, but with no upper limit
on the indices.  A typical element $A$ of $\Rs$ may then be expanded as
 \begin{equation}\label{GENEXPeq}
   \Scap{A} = \sum_{\lambda \in M_{\infty}} \Scap{A}_{\lambda}\beta_{\lambda}
 \end{equation}
where each coefficient $\Scap{A}_{\mu}$ is a real number. For each
$\lambda \in M_{\infty}$ there is a  \Defi{generator projection}
map
 \begin{equation}\label{GENPROJeq}
   P_{\lambda}:\Rs \to \Real ,
   \quad
   \Scap{A} \mapsto \Scap{A}_{\lambda},
 \end{equation}
a particular case of this being the augmentation map
 \begin{equation}\label{BODYeq}
   \epsilon:\Rs \to \Real ,
   \quad
   \Scap{A} \mapsto \Scap{A}_{\emptyset}.
 \end{equation}
Clearly $\Rs$ is an infinite-dimensional vector space. However our
use of $\Rs$ is purely algebraic, so that we  do not need to equip
it with a norm, and the following notion of convergence is
sufficient:

 \begin{Def}\label{CONVdef}
 { \rm
   A sequence $\Scap{X}_k, k= 1, \dots$ of elements of $\Rs$ is said
  converge to the limit $\Scap{X}$ in $\Rs$ if each coefficient
   $\Scap{X}_{k,\lambda}$ in the expansion \break
   $\Scap{X}_{k} = \sum_{\lambda \in M_{\infty}}
     \Scap{X}_{k,\lambda} \beta_{\lambda}$ converges to the coefficient
     $\Scap{X}_{\lambda}$ in the expansion \break
    $\Scap{X} = \sum_{\lambda \in M_{\infty}}
     \Scap{X}_{\lambda} \beta_{\lambda}$.
 }\end{Def}

The Grassmann algebra $\Rs$ is used to build $(m,n)$-dimensional
superspace $\Rstt{m}{n}$ in the following way:
 \begin{Def}
 { \rm
   $(m,n)$-dimensional \Defi{superspace} is the space
 \begin{equation}
   \Rstt{m}{n} = \underbrace{\Rs{}_0\times\dots\times \Rs{}_0}_{m {\rm\ copies}}
  \times \underbrace{\Rs{}_1\times\dots\times \Rs{}_1}_{n {\rm\ copies}}.
 \end{equation}
 }\end{Def}
A typical element of $\Rstt{m}{n}$ is written $(x; \xi)$ or
 $(x^1, \dots, x^m; \xi^1, \dots, \xi^n)$, where the convention is
 used that lower case Latin letters represent even objects and lower
 case Greek letters represent odd objects, while small capitals are used
 for objects of mixed or unspecified parity.

Consider first functions with domain the space $\Rstt{0}{n}$ of which a
typical element is
  $\xi := (\xi^{1},\ldots,\xi^{n})$, that is, functions of purely
anticommuting variables.  (For simplicity it will be assumed that
$n$ is an even number.) We will consider functions on this space
which are multinomials in  the anticommuting variables. Using the
multi-index notation introduced above, a multinomial function
$\Scap{F}$ may then be expressed in the standard form
 \begin{eqnarray}
  \Scap{F}:\Rstt{0}{n} & \longrightarrow & \Rs \nonumber \\
  (\xi^{1},\dots,\xi^{n}) & \mapsto &
  \sum_{\mu \in M_{n}} \Scap{F}_{\mu} \xi^{\mu} \label{susmfn}
 \end{eqnarray}
where the coefficients $\Scap{F}_{\mu}$ are real  numbers. Such
functions will be known (anticipating the terminology for
functions of both odd and even variables) as \Defi{supersmooth}.
More general supersmooth functions, with the coefficients
$\Scap{F}_{\mu}$ taking values in $\Comp$, $\Rs$, or some other
algebra are also possible.

Differentiation of supersmooth functions of anticommuting
variables is defined by linearity together with the rule
 \begin{equation}
  \frac{\partial \xi^{\mu}}{\partial \xi^{j}} =
  \left\{
  \begin{array}{cl}
 (-1)^{{k}-1} \xi^{\mu_{1}} \ldots   \widehat{\xi^{{k}}} \ldots \xi^{\mu_{|\mu|}},
   & \Mbox{ if $j=\mu_{{k}}$ for some ${k}$, $1 \leq {k} \leq |\mu|$,} \\
 0 & \Mbox{otherwise,}
  \end{array}
  \right.
 \label{graddiff}
 \end{equation}
where  the caret\ {}$\,{\widehat{}}\,${}\ indicates an omitted
factor.

Integration of functions of purely anticommuting variables is
defined algebraically by the Berezin rule
\cite{Martin,Berezi1966}:
 \begin{equation}\label{BINTeq}
  \Bint d^n\xi \, \Scap{F}(\xi) = \Scap{F}_{1 \ldots n},
 \end{equation}
where $\Scap{F}(\xi)= \sum_{\mu \in M_{n}} \Scap{F}_{\mu}
\xi^{\mu}$ as in (\ref{susmfn}), so that $\Scap{F}_{1 \ldots n}$
is the coefficient of the highest order term.

This integral can be used to define a  Fourier transform with
useful properties.  If $\Scap{F}$ is a supersmooth function on
$\Rstt{0}{n}$, then the Fourier transform $\Ft{\Scap{F}}$ of
$\Scap{F}$ is defined by
 \begin{equation}
   \Ft{\Scap{F}} (\rho)
   = \Bint \Df^n \xi \, \Scap{F}(\xi) \exp i  \rho.\xi
 \end{equation}
where $\rho.\xi = \sum_{i=1 \dots n} \rho^i \xi^i$. A simple calculation
establishes the Fourier inversion theorem
 \begin{equation}\label{FITeq}
   \Ft{\Ft{\Scap{F}}}= \Scap{F}.
 \end{equation}
Any linear operator $K$ on the space of supersmooth functions of
purely anticommuting variables has integral kernel $\Ker{K}$
taking
 $\Rstt{0}{n} \times \Rstt{0}{n} $ into $\Rs$ defined
by
 \begin{equation}\label{IKeq}
  K \Scap{F}(\xi) = \Bint d^n\theta \, \Ker{K}(\xi,\theta) \Scap{F}(\theta).
 \end{equation}

Using the Fourier inversion theorem we see that
 \begin{eqnarray}
   \delta(\xi,\theta)
   &=& \Bint \Df^n \rho \, \Expi{- \rho . (\xi - \theta)}\End
   &=& \prod_{i=1}^{n}  (\xi^i-\theta^i)
 \label{DELFeq}
 \end{eqnarray}
 is the kernel of the identity operator, that is
 \begin{equation}
   \int \Df^n \theta \delta(\xi,\theta) \Scap{F}(\theta)
   = \Scap{F}(\xi).
 \end{equation}

More generally, if $K$ is a differential operator acting on
supersmooth functions on $\Rstt{0}{n}$, then
 \begin{equation}
   \Ker{K} (\xi, \theta) = K_{\xi} \delta (\xi, \theta)
 \end{equation}
where the subscript $\xi$ indicates that derivatives are taken
with respect to this variable.

The link between anticommuting variables and Clifford algebras, which
underpins the various constructions in this paper, arises from the
operators
 \begin{equation}\label{CLIFFORDOPSeq}
  \psi^i = \xi^i + \Dbd{\xi^i}, \qquad i=1, \dots, n
 \end{equation}
which satisfy the anticommutation relations
 \begin{equation}\label{ANTICOMeq}
 \Com{\psi^i}{\psi^j} =  \psi^i \psi^j + \psi^j \psi^i = \delta^{ij}.
 \end{equation}
(In the super algebra context, the commutator of two operators $A$ and
$B$ is defined by $\Com{A}{B} = AB - \Cfac{A}{B} B A$.)

If $\mu$ is a multi-index in $M_n$ then
 \begin{equation}\label{KERNELeq}
   \Ker{\psi^{\mu}}(\xi, \theta)
   = \Bint \Df^n \rho (\xi+i \rho)^{\mu}
                                \Expi{-\rho . (\xi - \theta)},
  \end{equation}
a fact which leads to the fermionic Feynman-Kac formulae exploited in
this paper.

In the case of operators acting on a graded algebra a useful and natural
quantity is the supertrace. If
$\Wi$ is the operator on the algebra in question which acts as $1$ on
even elements and $-1$ on odd elements, the supertrace is of an operator
$K$ is defined to be the trace of the operator $\Wi K$. It may readily be
shown that integration of the kernel of $K$ at coincident points gives
the supertrace, that is
 \begin{equation} \label{STRFeq}
  \Str K = \Bint d^n\xi \, \Ker{K}(\xi,\xi)
 \end{equation}
while the standard trace may be obtained from the formula
  \begin{equation} \label{TRFeq}
  \Tr K = \Bint d^n\xi \, \Ker{K}(\xi,-\xi).
 \end{equation}
More general supertraces also occur, where $\gamma$ is replaced by
some other involution.

In order to extend the notion of supersmooth to functions on the
more general superspace $\Rstt{m}{n}$, we must take note of the
fact that an even Grassmann variable is not simply a real or
complex variable. However the necessary class of functions can be
captured by defining supersmooth functions on $\Rstt{m}{0}$ as
extensions by Taylor expansion from smooth functions on $\Real^m$.

 \begin{Def}
 { \rm
   The function $\Scap{F}: \Rstt{m}{0} \to \Rs$ is said to be
   \Defi{supersmooth} if there exists a function $\tilde{\Scap{F}}:\Real^m \to
   \Rs$ whose combination with each generator projection (c.f.
   (\ref{GENPROJeq})) is smooth, such that
 \begin{eqnarray}
  \Scap{F}(x^1, \dots, x^m)&=& \tilde{\Scap{F}}(\epsilon(x))
   + \sum_{i=1}^{n} (x^i- \epsilon(x^i)\One)
   \frac{\partial  \tilde{\Scap{F}}}{\partial x^i}(\epsilon(x)) \End
   && +\sum_{i,j=1} (x^i- \epsilon(x^i)\One)(x^j- \epsilon(x^j)\One)
   \frac{\partial^2  \tilde{\Scap{F}}}{\partial x^i \partial x^j}
   (\epsilon(x)) \dots \, .
 \end{eqnarray}
(Although this Taylor series will be infinite, it gives well
defined coefficients for each $\beta_{\lambda}$ in the expansion
(\ref{GENEXPeq}), so that the value of $\Scap{F}$ is a
well-defined element of $\Rs$.)
  }\end{Def}
 A supersmooth function on the general superspace $\Rstt{m}{n}$
 can now be defined.
 \begin{Def}\label{SSFUNCdef}
 { \rm
   A function $\Scap{F}: \Rstt{m}{n} \to \Rs$ is said to be supersmooth
   if there exist supersmooth functions $\Scap{F}_{\mu}, \mu \in M_n$ of
   $\Rstt{m}{0}$ into $\Rs$  such that
 \begin{equation}
   \Scap{F}(x, \xi) = \sum_{\mu\in M_{n}} \Scap{F}_{\mu} (x) \xi^{\mu}
 \end{equation}
for all $(x,\xi)$ in $\Rstt{m}{n}$.
 }\end{Def}

Integration of supersmooth functions is defined by a combination of
Berezin integration and conventional Riemann integration.
 \begin{Def} \label{SUPERINTEGRALdef} {\rm
 If $\Scap{F}:\Rstt{m}{n} \to \Rs$ is supersmooth and $V \subset \Real^m$,
 then the integral of $\Scap{F}$ over $V$ is defined to be
 \begin{equation}\label{SUPERINTEGRALeq}
  \Sint{V} d^mx \, d^n \xi \, \Scap{F}(x, \xi)
  = \int_{V} d^mx \, \left( \Bint d^n\xi \, \Scap{F}(x, \xi) \right).
 \end{equation}
 }
 \end{Def}
Using the Berezinian, which is the superdeterminant of the matrix of
partial derivatives, a change of variable rule may be obtained which is
valid for functions of compact support.

Up to this point we have used the prefix super merely to indicate the
presence of an anticommuting extension of some classical commuting
object. The concept of supersymmetry involves the further feature that
the Hamiltonian (or, geometrically, the Laplacian) of the system is the
square of an odd operator known as the supercharge, or the sum of squares
of several supercharges. Many examples of this are given in the following
section on supersymmetric quantum mechanics.  The corresponding time
evolution may also have a square root, which will be defined by
introducing an odd parameter $\tau$ in conjunction with the usual time
parameter $t$, and defining some rather special objects on the
$(1,1)$-dimensional superspace parametrised by $(t;\tau)$. The starting
point is the superderivative
 $D\Subtt = \Dbd{\tau} + \tau \, \Dbd t$ acting on functions
$\Fsc(t; \tau)$ on the superspace  $\Rstt11$. Since $(\Dbd{\tau})^2=0$,
this has the property that
 \begin{equation}
  (D\Subtt)^2 = \Dbd{t},
 \end{equation}
so that $D\Subtt$ is a square root of the generator of time
translations.  It is also possible to introduce a notion of
integration between even and odd limits of a function $\Fsc$ on
$\Rstt11$ in the following way:
 \begin{equation}\label{SUPERINTLIMeq}
  \Intzztt \Dss \Fsc(s;\sigma) =
  \Bint d\sigma \int_0^{t+\sigma\tau} ds \Fsc(s;\sigma).
 \end{equation}
It may then be shown by direct calculation that this integral
provides a square root of the fundamental theorem of calculus in
the sense that that
 \begin{equation}\label{SFUNDTHECALCeq}
  \Intzztt \Dss D\Subss \Fsc(s;\sigma)
  = \Fsc(t;\tau) - \Fsc(0;0).
  \end{equation}
If we now introduce a superpath $X:\Rstt11 \to \Rstt{p}{q}$
 and let
 \begin{equation}
 DX^i= \Dss D\Subss X^i(s, \sigma), i=1, \dots p+q,
 \end{equation}
then by using the chain rule for derivatives together with
(\ref{SFUNDTHECALCeq}) the integral along $X$ of the gradient of a
function $\Scap{G}$ on $\Rstt{p}{q}$ can be expressed as the
difference between the values of $\Scap{G}$ at the endpoints of
the super path:
\begin{equation}\label{INTALONGSUPERPATHeq}
  \Intzztt DX^i \Ds{i} \Scap{G}(X(s;\sigma)) =
  \Scap{G}(X(t;\tau)) - \Scap{G}(X(0;0)).
\end{equation}
(The usual summation convention is applied for repeated indices.) In
section \ref{DIRACsec} a stochastic version of this result is established
and applied to the Dirac operator.

In many applications of anticommuting variables the simple superspace
$\Rstt{m}{n}$ must be replaced by a more general supermanifold.  In this
article it will be sufficient to consider  supermanifolds  constructed in
a standard way from the data of a smooth vector bundle $E$ over a smooth
manifold $M$. (A theorem of Batchelor
\cite{Batche1979} shows that all smooth supermanifolds may be
obtained in this way.) The idea of the construction is   to patch
together local pieces of the supermanifold using the change of
coordinate functions of the manifold and the transition functions
of the bundle.  A careful definition would need an excursion into
supermanifold theory \cite{GTSM}; here it will suffice to define
change of coordinate functions, a full description of the patching
construction may be found in \cite{ERICE}.

Suppose that $M$ has dimension $m$ and $E$ has dimension $n$; the
supermanifold $\Sm{M}{E}$ is then $(m,n)$-dimensional.  If
 $\{ U_{\alpha} | \alpha \in \Lambda  \}$ is an open cover
of $M$ by sets which are both coordinate neighbourhoods of $M$ and
local trivialization neighbourhoods of $E$, then for each $\alpha$
in $\Lambda$ there are local coordinates
 $\Srow{x_{\alpha}}{m}{\xi_{\alpha}}{n}$ for the supermanifold;
change of coordinates on any overlap between charts is defined in
terms of the coordinate maps $\phi_{\alpha}:U_{\alpha} \to
\Real^m$ of $M$ and the vector bundle transition functions
 $g_{\alpha\beta}:U_{\alpha} \cap U_{\beta} \to \Mbox{GL}(n | \Real)$
 by
  \begin{eqnarray}
  x^i_{\beta}(x_{\alpha};\xi_{\alpha})&=&
   \phi^i_{\beta}(x_{\alpha}) \qquad i=1, \dots, m  \End
  \xi^j_{\beta}(x_{\alpha};\xi_{\alpha})&=&
   \sum_{k=1}^n g\,{}_{\alpha\beta}{}^j{}_k(x_{\alpha}) \xi_{\alpha}^k
   \qquad j=1, \dots, n.
 \end{eqnarray}

A particular example of this construction is the supermanifold
$\Sm{M}{TM}$ obtained from the tangent bundle of a manifold $m$.
Explicitly, if $M$ has dimension $m$, then $\Sm{M}{TM}$ has dimension
$(m,m)$, and local coordinates $\Srow{x_{\alpha}}{m}{\xi_{\alpha}}{m}$
which change according to the rule
 \begin{eqnarray}
  x^i_{\beta}(x_{\alpha};\xi_{\alpha})&=&
   \phi^i_{\beta}(x_{\alpha}) \qquad i=1, \dots, m  \End
  \xi^j_{\beta}(x_{\alpha};\xi_{\alpha})&=&
   \sum_{k=1}^{m} \Dbdf{x_{\beta}^j}{x_{\alpha}^k}(x_{\alpha})\, \xi_{\alpha}^k
   \qquad j=1, \dots, m.
 \end{eqnarray}
Supersmooth functions on this supermanifold are then naturally identified
with forms on $M$; in local coordinates this identification may be
expressed as
 \begin{equation}\label{FORMFUNCTIONeq}
  \Sum{\mu \in M_m}{} f_{\mu}(x) \xi^{\mu} \leftrightarrow
\Sum{\mu \in M_m}{} f_{\mu}(x) dx^{\mu},
 \end{equation}
so that Berezin integration on the supermanifold corresponds to the
standard integration of top forms on the manifold and the exterior
derivative takes the form $d = \xi^{i}\Dbd{x^i}$.

For geometric applications of superspace path integration, the
significant supermanifold is
$\Som$, the underlying manifold being the orthonormal frame bundle $O(M)$
of a Riemannian manifold
$(M,g)$, and the vector bundle over $O(M)$ being the bundle induced by
projection of
$O(M)$ onto $M$ of the product of the tangent bundle of $M$ and a bundle
$E$ over $M$. There is a natural definition of Brownian motion on this
supermanifold,  based on the Brownian motion on manifolds defined by
Elworthy \cite{Elwort} and by Ikeda and Watanabe \cite{IkeWat}, whose
construction is described in section \ref{SWMsec}.

A further supermanifold with geometric applications is the supermanifold
$\Sm{M}{T(M) \otimes E}$, where $E$ is an $n$-dimensional Hermitian vector
bundle over $m$; if one takes local coordinates
$(x^1, \dots,x^m; \xi^1, \dots, \xi^m, \eta^1, \dots, \eta^n)$, then
supersmooth functions which are linear in the $\eta$ variables correspond
to forms on $M$ twisted by $E$.  Given a metric $g$ on $M$ and a
connection $A$ for $E$, the Hodge-de Rham operator takes the form
  \begin{equation}\label{HODGEDERHAMeq}
  d + \delta =
\psi^{i} \left(  \Dbd{x^i} - \Chri{i}{j}{k} \xi^j \Dbd{\xi^k}
 - A_{i \, r}{}^s \eta^r \Dbd{\eta^s}\right)
 \end{equation}
where $\Chri{i}{j}{k}$ are the Christoffel symbols for the metric
$g$ and the Clifford algebra operators $\psi^i$
are now adapted to curved space, taking the form
$\psi^i = \xi^i + g^{ij}(x) \Dbd{\xi^j}$. The corresponding
Weitzenbock formula is then
 \begin{equation}\label{WFlem}
 -2(d + \delta)^2 =
 B - R_i^j(x) \xi^i \Dbd{\xi^j} -
 \Half R_{ki}{}^{jl}(x) \xi^{i}\xi^{k}\Dbd{\xi^{j}}\Dbd{\xi^{k}}
+ \Frac14 [\psi^i, \psi^j] {F}_{ij \, r}{}^s(x) \eta^r
\Dbd{\eta^s}
 \end{equation}
where $R_{ki}{}^{jl}$ are the components of the curvature of $g$, $F_{ij
\, r}{}^{s}$ are the components of the curvature of $A$ and $B$ is the
twisted Bochner Laplacian
 \begin{equation}\label{BOCNERLAPLACeq}
  B= g^{ij}\left( D_{i} D_{j} - \Chri{i}{j}{k} D_{k} \right)
 \end{equation}
with $D_i = \Dbd{x^i}- \Chri{i}{j}{k} \xi^j \Dbd{\xi^k}
                     - A_{i \, r}{}^{s} \eta^r \Dbd{\eta^s}$.
(A proof of this result, generalising the proof given in \cite{Simetal},
may be found in \cite{SCSONE}.)

The Dirac operator for a spin manifold $M$ of even dimension $m$
may also be represented as a differential operator on a
supermanifold. In this case the supermanifold is constructed from
the vector bundle $E_{O(M)}$ associated to the bundle of
orthonormal frames of $M$ via the vector representation of
$SO(m)$. The supermanifold $\Spinsm$ has local coordinates
$(x^i;\eta^a)$, $a,i=1, \dots, m$, using the convention that
coordinate indices are from the middle of the alphabet while
orthonormal frame indices are from the beginning of the alphabet.

Defining operators $\psi^a = \eta^a + \Dbd{\eta^a}$ we see that
$\psi^a \psi^b + \psi^b \psi^a = 2 \delta^{ab}$, so that the
supersmooth functions over a point of $M$ define a
$2^m$-dimensional representation of the Clifford  algebra on
$\Real^m$. The Dirac representation can obtained by considering
supersmooth functions with values in $\Rs \otimes \Comp$, and
restricting to the $2^{m/2}$-dimensional subspace of functions
which satisfy the $n$ conditions
 \begin{equation}
 \Psib^{2r-1}\Psib^{2r} \,f= i f, \quad r=1, \dots, m/2,
 \label{DCeq}\end{equation}
 with $\Psib^a = \eta^a - \Dbd{\eta^a}$.

 Functions satisfying these conditions will be referred to as Dirac
functions. It is also useful to define the projection operator $P$
which projects arbitrary functions onto Dirac functions by setting
 \begin{equation}
 P= P_1 \dots P_{n/2}
 \end{equation}
where, for each $r=1,\dots,n/2$, $P_r$ is the operator which
satisfies
 \begin{eqnarray}
 \lefteqn{P_r\left(g(\eta^1,\dots,\eta^{2r-2})(a +b\eta^{2r-1} + c\eta^{2r} +
 d\eta^{2r-1} \eta^{2r}) h(\eta^{2r+1},\dots, \eta^n)\right)=} \End
 && g(\eta^1,\dots,\eta^{2r-2})\left(\Frac{a-id}{2}(1-i\eta^{2r-1}\eta^{2r})
 + \Frac{b+ic}{2}(\eta^{2r-1} -i\eta^{2r})\right)
 h(\eta^{2r+1},\dots, \eta^n).\End
 \end{eqnarray}
  \section{Fermionic and supersymmetric quantum\hfil\break mechanics}
\label{SUSYQMsec}
In this section we describe the quantum mechanical models whose heuristic
path integral quantization is constructed rigorously in this survey. We
begin with a purely fermionic model, and investigate various
supersymmetric models, starting in flat space and then moving to curved
space where we develop the models used section \ref{ASIsec} for the
supersymmetric proof of the index theorem  and in section \ref{MTTsec}
for the study of Morse theory.

In the canonical quantization of $n$-dimensional particle mechanics, the
classical observables $p_i$ (momentum) and $x^i$ (position) are replaced
by quantum operators which satisfy the canonical commutation relations
$[x^i,p_j]=i \delta^i_j$. The standard representation is the Schr\"odinger
representation, with $x^i, i=1, \dots,m$ realised as the multiplication
operator on $\Ltr{m}$, and $p_i= -i\Dbd{x^i}$.  For fermionic operators
$\psi^i$ the canonical anticommutation relations can be represented as in
an analogous manner on functions of anticommuting variables $\xi^i$ by
setting $\psi^i = \xi^i + \Dbd{\xi^i}$, so that
 $\Com{\psi^i}{\psi^j} = \delta^{ij}$.

When fermions and bosons are both present, wave functions are functions
$\Scap{F}(x,\xi)$, with time evolution determined as usual by the Schr\"odinger
equation $ i \Dbdf{\Scap{F}}{t} =  H \Scap{F}$ (or, in Euclidean time,
$ \Dbdf{\Scap{F}}{t} = -  H \Scap{F}$), where units are
used in which Plank's constant $\hbar$ is equal to $1$.  In flat space
the standard Hamiltonians take the form
 $H = \frac{1}{2m} p_i p_i + V(x, \xi)$. The free Hamiltonian, on which the super
Wiener measure developed in sections \ref{GPSsec} and \ref{SWMsec} is
based, is
 $H_{0}= \frac{1}{2m} p_i p_i$; the fermionic contribution to the free
Hamiltonian is zero. (Generally we will use $m=1$.)

The defining feature of a supersymmetric theory is that the Hamiltonian
 $H$ has the form $H= \Half \Com{Q}{Q} = Q^2$ (or, more generally,
 $H=\Half \sum \Com{Q_i}{Q_i} = \sum Q_i^2$) where $Q$ (or $Q_i$) is an
odd `supercharge'.  The Lagrangian for such a theory is symmetric under a
group of transformations which includes transformations of fermions into
bosons and vice versa.  The simplest example of a supersymmetric
Hamiltonian is the free Hamiltonian $H_{0} = Q_{0}^2$ with
 $Q_0= \psi^i p_i$.  In sections
\ref{ASIsec}, \ref{DIRACsec} and \ref{MTTsec} geometric examples of
supersymmetric Hamiltonians are used.

\section{Grassmann probability spaces and \hfil\break fermionic Wiener
measure}
 \label{GPSsec}
In this section we begin by defining a notion of Grassmann probability
space and random variable, based on the standard finite-dimensional
Berezin integral (\ref{BINTeq}) for functions of anticommuting variables.
The Berezin integral is essentially algebraic; it is neither the limit of
a sum nor an antiderivative, and lacks the positivity properties
necessary for the use of standard measure theoretic techniques. The
approach to defining an analogue of probability measure taken here, which
was first developed in \cite{GBM}, is to use finite dimensional marginal
distributions and  Kolmogorov consistency conditions.

A particular example, Grassmann Wiener space, together with the
corresponding fermionic Brownian motion, is then constructed and its
relationship to the heat kernel of fermionic Hamiltonians is
demonstrated.

 \begin{Def}
 { \rm
An $n$\Defi{-Grassmann probability space of weight }$w$,  where
$w$ is an element of $\Rs$, consists of
 \begin{enumerate}
  \item a set $A$;
  \item for each finite subset $B$ of $A$ a supersmooth function
  $\Scap{F}_{B}: (\Rstt{0}{n})^{B} \to \Rs$ such that if
 $B=\{b_1, \dots,b_N \}$ and $B'=\{b_1, \dots,b_{N-1} \}$ then
 \begin{enumerate}
  \item
   \begin{equation}
     \Bint \Df^{n} \theta_{b_1} \dots  \Df^{n} \theta_{b_N} \,
     \Scap{F}_{B}(\theta_{b_1}, \dots \theta_{b_N}) = w,
   \end{equation}
  \item
 \begin{equation}
     \Bint \Df^{n} \theta_{b_N} \, \Scap{F}_{B}(\theta_{b_1}, \dots, \theta_{b_N}) =
     \Scap{F}_{B'}(\theta_{b_1}, \dots, \theta_{b_{N-1}}).
   \end{equation}
  \end{enumerate}
 \end{enumerate}
 Such a space will be denoted
 $\left( \Rstt{(0,n)}{A}, \{\Scap{F}_{B} \} \right)$.
 }\end{Def}
Having built Kolmogorov consistency conditions into the
definition, a notion of Grassmann random variable can be defined
by a limiting process. The definition of such random variables is
rather cumbersome because the generalised measure we are using
does not have the usual positivity properties.

 \begin{Def}\label{GRVdef}
 { \rm
   Suppose that $\Fclass$ is a class of functions on
   $\Rstt{r}{s}$. Then an $(r,s)$-dimens\-ion\-al Grassmann
   random variable of class $\Fclass$ on a
   Grassmann probability space
   $\left( \Rstt{(0,n)}{A}, \{\Scap{F}_{B} \} \right)$
   consists of
 \begin{enumerate}
  \item a sequence $B_1= \{ b_{1,1}, \dots, b_{1, \Num{B_1}}\},
  B_2 =  \{ b_{2,1}, \dots, b_{2, \Num{B_2}}\},
   \dots$ of finite subsets of $A$,
   where $\Num{B_k}$ denotes the number of elements in the set $B_k$;
  \item a sequence  $\Scap{G}_1,\Scap{G}_2, \dots$ of supersmooth
  functions, with $\Scap{G}_k : (\Rstt{0}{n})^{B_k} \to \Rstt{r}{s}$, such
  that for each function $\Scap{H}$ in $\Fclass$ the sequence
 \begin{eqnarray}
   \lefteqn{I_{k}\Scap{H} = } \End
 &&  \Bint \Df^{n} \theta_{b_{k,1}} \dots  \Df^{n} \theta_{b_{k,\Num{B_k}}} \,
     \Scap{F}_{B}(\theta_{b_{k,1}}, \dots \theta_{b_{k,\Num{B_k}}}) \End
 && \qquad \times    \Scap{H}( \Scap{G}_{k}(\theta_{b_{k,1}}, \dots \theta_{b_{k,\Num{B_k}}}))
 \end{eqnarray}
tends to a limit in $\Rs$ as $k \to \infty$.  (The notion of
convergence in $\Rs$ is given in Definition \ref{CONVdef}.)
 \end{enumerate}
 The limit of this sequence is the \Defi{Grassmann expectation} of $\Scap{F}(\Scap{G})$,
 and is denoted $\Exg{\Scap{F}(\Scap{G})}$.
 }\end{Def}

 We now construct fermionic Wiener measure, which is the key
construction underpinning the probabilistic methods described in this
survey. The  finite dimensional marginal distributions which determine
this measure are built from the heat kernel of the free fermionic
Hamiltonian. Since this Hamiltonian is zero, the measure is in fact built
from Grassmann delta functions (\ref{DELFeq}). In order to obtain a
Feynman-Kac formula which can handle differential operators of quite a
general class, the measure is over paths in phase space, that is, the
Fourier transform variables are not integrated out before defining the
measure.

 \begin{Def}\label{FWMdef}
 { \rm
 \begin{enumerate}
  \item Let $A$ be the interval $[0, \infty)$.
  Then $n$-dimensional fermionic Wiener space is
  defined to be the $2n$-Grassmann probability space
  $(\Rstt{0}{2n})^{A}, \{ \Scap{F}_{B} \})$ such that, given
  $B= \{ t_1, \dots, t_N \} \subset A$ with
  $0 \leq t_1 < t_2 <  \dots <t_N $,
 \begin{eqnarray}
   \lefteqn{\Scap{F}_{B}(\theta_1,\rho_1, \dots, \theta_N, \rho_N)} \End
   & =& \Expi{- \rho_1 . \theta_1 - \rho_2 . (\theta_2 - \theta_1)
   - \dots - \rho_N.(\theta_N- \theta_{N-1})}
 \end{eqnarray}
where $\rho_k, \theta_k, k = 1 \dots N$ are in
 $\Rstt{0}{n}$. (Here we use $\rho_k, \theta_k$ rather than
 $\rho_{t_k}, \theta_{t_k}$ because the ordering of the elements of
 $B$ is specified.)
 \item The corresponding stochastic process
 $(\theta_t{},\rho_t{})$ is called
\Defi{fermionic Brownian motion}.
 \end{enumerate}
 }\end{Def}
 The following example illustrates this measure in
action.
 \begin{Exa}
 {\rm
 For any supersmooth function $\Scap{F}$ of $n$
anticommuting variables with
 $\Scap{F}(\xi) = \sum_{\mu \in M_n} \Scap{F}_{\mu} \xi^{\mu}$
 define the operator  $\Scap{F}(\psi)$  to be
 $\sum_{\mu \in M_n} \Scap{F}_{\mu} \psi^{\mu}$
 where $\psi^i$ are as before the Clifford algebra or fermionic operators
 $\psi^i = \xi^i + \Dbd{\xi^i}$.
   Let $\Scap{F}_1, \dots, \Scap{F}_N$ and $\Scap{G}$ be supersmooth functions
of $n$ anticommuting
variables and $t_1, \dots, t_N$ be real numbers with
 $0 <t_1 < \dots<t_N$.
The  action of the operator $K = \Scap{F}_1(\psi) \dots
\Scap{F}_N(\psi)$ on the function $\Scap{G}$ is then given by
 \begin{equation}
  K\,\Scap{G}(\xi^1, \dots,\xi^n)
= \Exgb{\Scap{F}_1(\theta_{t_1}+i \rho_{t_1}) \dots
\Scap{F}_N(\theta_{t_N} + i rho_{t_N})\Scap{G}(\theta_{t_N})}.
 \end{equation}
 This result follows from repeated application of (\ref{KERNELeq}) }.
 \end{Exa}
 We end this section with a lemma that is useful when estimating
fermionic integrals.

 \begin{Lem}\label{FEXlem}
   Let $\Scap{F}$ be a finitely defined random variable on $n$-dimensional
fermionic Wiener space, with
 \begin{eqnarray}
  \lefteqn{\Scap{F} (\theta_{t_1}, \rho_{t_1},\dots, \theta_{t_N},\rho_{t_N})=
 \sum_{\mu_1 \in M_n}\sum_{\nu_1 \in M_n} } \End
&& \dots \sum_{\mu_N \in M_n} \sum_{\nu_N \in M_n}
 F_{\mu_1 \nu_1 \dots \mu_N \nu_N}
 \theta_{t_1}^{\mu_1} \rho_{t_1}^{\nu_1} \dots
\theta_{t_N}^{\mu_N}\rho_{t_N}^{\nu_N},
 \end{eqnarray}
then
 \begin{equation}
  |\Exgb{\Scap{F}}| \leq
 \sum_{\mu_1 \in M_n}\sum_{\nu_1 \in M_n} \dots \sum_{\mu_N \in M_n} \sum_{\nu_N \in M_n}
 |F_{\mu_1 \nu_1 \dots \mu_N \nu_N}|.
 \end{equation}
\end{Lem}
\Oproof The result follows from the Berezin integration rule, together
with the fact that each term in the expansion of $\exp (-i\rho.\theta)$
occurs with coefficient of size exactly one.

 \section{Super Wiener measure and
 It\^o integration along super Brownian paths}
 \label{SWMsec}
In this section we begin by combining some ideas from conventional
measure theory with the anticommuting probability measures of the
preceding section to construct a notion of super probability
space, together with its associated random variables.  Again we
use finite dimensional marginal distributions as the basis of the
construction.
  \begin{Def}\label{SPSdef}
 { \rm
An $(m,n)$\Defi{-super probability space of weight }$w$,  where
$w$ is an element of $\Rs$, consists of
 \begin{enumerate}
  \item a set $A$;
  \item for each finite subset $B$ of $A$ a supersmooth function
  $\Scap{F}_{B}: (\Rstt{m}{n})^{B} \to \Rs$ such that if
 $B=\{b_1, \dots,b_N \}$ and $B'=\{b_1, \dots,b_{N-1} \}$ then
 \begin{enumerate}
  \item
   \begin{equation}
     \Sint{\Real^m} \Df^{m}_{b_1}x \Df^{n} \theta_{b_1}
     \dots  \Df^{m}x_{b_N} \Df^{n} \theta_{b_N} \,
     F_{B}(x_{b_1},\theta_{b_1}, \dots x_{b_N},\theta_{b_N}) = w,
   \end{equation}
  \item
 \begin{eqnarray}
  \Sint{\Real^m} \Df^{m}x_{b_N}\Df^{n} \theta_{b_N} \,
    \lefteqn{ \Scap{F}_{B}(x_{b_1},\theta_{b_1}, \dots, x_{b_N},\theta_{b_N})}
   \End
      &=&    \Scap{F}_{B'}(x_{b_1},\theta_{b_1}, \dots,
                 x_{b_N},\theta_{b_{N-1}}).
   \end{eqnarray}
  \end{enumerate}
 \end{enumerate}
 Such a space will be denoted
 $\left( \left(\Rstt{m}{n}\right)^{A}, \{\Scap{F}_{B} \} \right)$.
 }\end{Def}
Because of the presence of anticommuting variables this `super'
probability measure is not true probability measure, so that a
specific definition of random variables is required. As with the
definition of Grassmann random variables, Definition \ref{GRVdef},
an explicit limiting process is used.
 \begin{Def}\label{SRVdef}
 { \rm
   Suppose that $\Fclass$ is a class of functions on
   $\Rstt{r}{s}$. Then an $(r,s)$-dimens\-ion\-al super
   random variable of class $\Fclass$ on on a
   super probability space
   $\left( \left(\Rstt{m}{n}\right)^{A}, \{\Scap{F}_{B} \} \right)$
   consists of
 \begin{enumerate}
  \item a sequence  $B_1= \{ b_{1,1}, \dots, b_{1, \Num{B_1}}\},
  B_2 =  \{ b_{2,1}, \dots, b_{2, \Num{B_2}}\},
   \dots$  of finite subsets of $A$;
  \item a sequence  $\Scap{G}_1,\Scap{G}_2, \dots$ of supersmooth
  functions, with $\Scap{G}_k : (\Rstt{m}{n})^{B_k} \to \Rstt{r}{s}$ such
  that for each function $\Scap{H}$ in $\Fclass$ the sequence
 \begin{eqnarray}
   \lefteqn{I_{k}\Scap{H} =
    \Sint
   \Df^{m} x_{b_{k,1}}\Df^{n} \theta_{b_{k,1}} \dots
   \Df^{m} x_{b_{k,\Num{B_k}}}\Df^{n} \theta_{b_{k,\Num{B_k}}} }\End
    && \Scap{F}_{B}(x_{b_{k,1}},\theta_{b_{k,1}},
     \dots x_{b_{k,\Num{B_k}}}, \theta_{b_{k,\Num{B_k}}}) \End
    && \quad \times \Scap{H}( \Scap{G}_{k}(x_{b_{k,1}},\theta_{b_{k,1}},
     \dots x_{b_{k,\Num{B_k}}}, \theta_{b_{k,\Num{B_k}}}))
 \end{eqnarray}
tends to a limit in $\Rs$ as $k \to \infty$.  (The notion of
convergence in $\Rs$ is given in Definition \ref{CONVdef}.)
 \end{enumerate}
 The limit of this sequence is the \Defi{super expectation} of $\Scap{H}(\Scap{G})$,
 and is denoted $\Exsb{\Scap{H}(\Scap{G})}$.
 }\end{Def}

It is also useful to introduce the notion of stochastic process.
 \begin{Def} {\rm
Let $I$ be an interval of the real line.  Then a collection
 $\{ \Scap{W}_t{}|t \in I \}$
of $(r,s)$-dimensional super random variables on a super
probability space is called an $(r,s)$-dimensional
\Defi{stochastic process} if each each finite subcollection is a
super random variable of the appropriate dimension.
 }\end{Def}
The notion of different versions of a process, and of Gaussian
process, correspond directly to the conventional objects.
The stochastic process which will play a major role in the geometric
applications defined below is Brownian motion in superspace, which is
obtained by combining the fermionic Brownian paths constructed in section
\ref{GPSsec} with conventional paths. These paths are obtained from
super Wiener measure, which is essentially the product of standard Wiener
measure and the fermionic Wiener measure of Definition~\ref{FWMdef}.
 \begin{Def}\label{SWMdef}
 { \rm
 \begin{enumerate}
  \item $(m,2n)$-dimensional Super Wiener space is the super probability space
  $((\Rstt{m}{n})^A, \{ \Scap{F}_{B} \} )$ with $A$ the interval
  $[0,  \infty)$ and, given $B= \{t_1, \dots, t_N \} \subset A$ with
 $0 \leq t_1 < t_2 < \dots < t_N$,
 \begin{eqnarray}
   \Scap{F}_{B} (b_1, \theta_1, \rho_1, \dots ,b_N, \theta_N,  \rho_N)
   &=& P_{t_1}(b_1,\theta_1, \rho_1,0,0)
   P_{t_2-t_1}(b_2,\theta_2,\rho_2, b_1,\theta_1)
   \End
  \dots   && P_{t_N-t_{N-1}}(b_N,\theta_N,\rho_N, b_{N-1},\theta_{N-1})
 \end{eqnarray}
where
 \begin{equation}
   P_{t}(b,\theta, \rho, b',\theta')
   = \frac{1}{(2 \pi t)^{\frac{m}{2}}}\Exp{-\frac{(b-b')^2}{2t}}
    \Expi{- \rho.(\theta - \theta')}.
 \end{equation}
\item The corresponding $(m,2n)$-dimensional process
 $(b_t,  \theta_t, \rho_t)$
 is called \Defi{super Brownian motion}.
 \end{enumerate}
 }\end{Def}
The motivation for constructing this super stochastic process is to give
a rigorous mathematical formulation of the combined bosonic and fermionic
path integrals which play a key role in many of the applications of
supersymmetric quantum mechanics to geometry. Its use in solving
diffusion equations on certain geometric bundles is described section
\ref{BMSMsec} and applied in Sections
\ref{ASIsec}, \ref{DIRACsec} and \ref{MTTsec}.
\section{Stochastic calculus of super Brownian \hfil\break
 paths}
In this section some modest generalisations of stochastic calculus are
introduced, so that a useful notion can be developed of Brownian path on
the various supermanifolds we have introduced, and applied to various
geometric diffusions. We begin by defining the concepts of adapted
process and stochastic integrals in the following way:
 \begin{Def}\label{ADAPdef}{ \rm
A stochastic process $\{ \Scap{F}_s:s \in [0,t] \}$ on
$(m,n)$-dimensional super Wiener space   such that for each
 $s$ in $[0,t]$ the random variable $\Scap{F}_s$ is a function of
 $\{b_{s'},\theta_{s'}, \rho_{s'} |0 \leq s' \leq s\}$ is said to be
 \Defi{$[0,t]$-adapted}.   (The time interval, $[0,t]$,
 may be omitted when the context makes it clear.)
 }
 \end{Def}
 As in the classical case, it can then be shown by
direct calculation that if $ \Scap{F}_s$ is a $[0,t]$-adapted process and
$0 \leq s_1 \leq s_2 \leq s_3 \leq t$ then
 \begin{eqnarray}\label{INDeq}
  \Exsb{\Scap{F}_{s_1} \left(b^i_{s_3}-b^i_{s_2}\right)}
  &=& 0 \End
  \Mbox{and} \qquad
 \Exsb{\Scap{F}_{s_1}
 \left(b^i_{s_3}-b^i_{s_2}\right)\left(b^j_{s_3}-b^j_{s_2}\right)}
  &=& \Half \Exsb{\Scap{F}_{s_1}} (s_3 -s_2),
 \end{eqnarray}
so that the commuting part of the super Brownian motion will contribute
to results in stochastic calculus in the normal way. There is no analogue
of It\^o integration along odd Brownian paths because the measure does
not lead to increments which are of order any positive power of $\delta
t$; however, when applying super Brownian paths to diffusions, the use of
$d\theta$ is unnecessary, since the presence of the paths $\rho_t$ enables one to
obtain a Feynman-Kac formula for fermionic differential operators
directly rather than by application of It\^o calculus.

As a result, it is sufficient to consider integrals with respect
to time and with respect to even Brownian paths. The definitions,
which are rather cumbersome, are given here for completeness.  It
will emerge below that they do lead to natural analogues of the
results in stochastic calculus which are important in applications
to diffusions. The integral $\Intzt \Fsc_s ds$ of an adapted
process $\Fsc_s$ with respect to time will now be defined.
 \begin{Def} \label{TIdef}{\rm
Let $\Fsc_s$ be a $[0,t]$-adapted process on $(m,2n)$-dimensional
super Wiener space. Suppose that for each $s$ in $[0,t]$, $\Fsc_s$
corresponds, as in definition \ref{SRVdef}, to the sequence of
pairs of subsets and functions
 $\left(J_{s,M}, \Fsc_{s,M}\right)$,\break $M=1,2,\dots$ with
 $J_M= \cup_{r=1}^{2^M-1} J_{t_r,M}$ where, for $r=1, \dots 2^M-1$,
 $t_r= \frac{rt}{2^M}$. Also define functions $K_M$ on
 $(\Rstt{m}{n})^{J_M}$ by
\begin{equation}
  K_M = \Sum{r=1}{2^M-1} \Fsc_{t_r, M} \, \frac{t}{2^M}.
\end{equation}
Then, if the appropriate limits exist as $M$ tends to infinity,
the sequence $(J_M,K_M)$ defines a super random variable which is
denoted $\Intzt \Fsc_s \, ds$.
 } \end{Def}
The other stochastic integral needed is the integral along even
Brownian paths:
 \begin{Def}\label{IIdef}{\rm
With the notation of previous definition, and with $\Fsc_s^a$ an
$m$-dimensional super random variable, for each $M=1,2, \dots$,
define the functions $\Scap{\Scap{G}}_M$ on $(\Rstt{m}{2n})^{J_M}$
by
 \begin{equation}
  \Scap{G}_M   = \Sum{a=1}{m} \Sum{r=1}{2^M-1}
                      \Scap{F}_{t_r,M}^a  (x^{r+1,a}-x^{r,a}).
 \end{equation}
Then, if the appropriate limits exist as $M$ tends to infinity,
the sequence \break $\left( J_M,\Scap{G}_M \right)$ defines a
super random variable which is denoted
 $\Intzt \Fsc_s^a \, db_s^a$.
 }\end{Def}
A key result in the application of super Brownian paths to the study of
diffusions on supermanifolds is the following It\^o formula:
 \begin{The}\label{ITOFORMULAthe}
   Suppose that ${a}_t$ is a $p$-dimensional stochastic integral on
$m$-dimensional Wiener space with
 \begin{eqnarray}
 \lefteqn{  x^i_t - x^i_0 = \Intzt
 \sum_{\mu\in M_n, \nu \in M_n}g^i_{\mu\nu,s} \theta_s^{\mu} \rho_s^{\nu}\,
ds} \End
 && +  \Intzt
 \Sum{a=1}{m} \sum_{\mu\in M_n, \nu \in M_n}g^i_{\mu\nu,s}
 h^i_{\mu\nu \, a\,s} \theta_s^{\mu} \rho_s^{\nu}\,
 db_s^a,
\,\, i=1, \dots, p
 \end{eqnarray}
where for $i=, \dots, p$ and $a = 1, \dots m$ the processes
$g^i_{\mu\nu,t}$ and
$h^i_{\mu\nu\, a\,t}$ are adapted processes on $m$-dimensional
Wiener space with
 \[
  \Excb{ \Intzt |h^i_{\mu\nu s}|^2 ds } < \infty \quad \mbox{and} \quad
  \Excb{ \Intzt |g^i_{\mu\nu \, a\,s}| ds } < \infty. \]
 (Here $\Exc$ denotes expectation with respect to
classical Wiener measure.)  Also suppose that
 $F:\Rstt{p}{2n} \to \Rs$ is supersmooth with compact support. Let
$\partial_i, i= 1, \dots, p$ denote differentiation with respect to the
$i^{\rm th}$ even variable.
 Then if $0 <t' < t < \infty$,
 \begin{eqnarray}\label{ITOFORMULAeq}
   \lefteqn{ F({x}_t,\theta_t, \rho_t) - F({x}_{t'}, \theta_{t'}, \rho_{t'})
     \Muq
  \Intzt \Sum{i=1}{p} \partial_i F({x}_s,\theta_s, \rho_s)
 (h^i_s\,ds + \Sum{a=1}{m} g^i_{a\,s} \, db^a_s) }
 \hspace{11em}\End
 && + \Intzt \Sum{i=1}{p}\Sum{j=1}{p}\Sum{a=1}{m}
 \Half\partial_i\partial_j F({x}_s,\theta_s, \rho_s)
 g^i_{a\,s} g^j_{a\,s} \, ds.
\End
 \end{eqnarray}
 \end{The}
\Oproof
 \begin{equation}
  F({x}_t,\theta_t, \rho_t) - F({x}_{t'}, \theta_{t'}, \rho_{t'})
 = \Sum{r=0}{N} \Delta_r \, F
 \end{equation}
where $\Delta_r \, F =F({x}_{t_r},\theta_{t_r}, \rho_{t_r}) -
 F({x}_{t_{r-1}}, \theta_{t_{r-1}}, \rho_{t_{r-1}})$ with
 $t_r= rt/N$, \break $r= 0, \dots, N$.
Now
 \begin{eqnarray}
  \lefteqn{ F({x}_{t_r},\theta_{t_r}, \rho_{t_r}) -
 F({x}_{t_{r-1}}, \theta_{t_{r-1}}, \rho_{t_{r-1}})} \End
&=& F({x}_{t_r},\theta_{t_r}, \rho_{t_r}) -
 F({x}_{t_{r-1}}, \theta_{t_{r}}, \rho_{t_{r}}) \End
&& + F({x}_{t_{r-1}},\theta_{t_r}, \rho_{t_r}) -
 F({x}_{t_{r-1}}, \theta_{t_{r-1}}, \rho_{t_{r-1}})
 \end{eqnarray}
and hence, since
 \begin{equation}
 \Exsb{ F({x}_{t_{r-1}},\theta_{t_r}, \rho_{t_r}) -
 F({x}_{t_{r-1}}, \theta_{t_{r-1}}, \rho_{t_{r-1}})} =0,
 \end{equation}
as may be seen directly from the definition of fermionic Wiener measure,
\begin{equation}
   F({x}_{t_r},\theta_{t_r}, \rho_{t_r}) -
 F({x}_{t_{r-1}}, \theta_{t_{r-1}}, \rho_{t_{r-1}})
\Muq F({x}_{t_r},\theta_{t_r}, \rho_{t_r}) -
 F({x}_{t_{r-1}}, \theta_{t_{r}}, \rho_{t_{r}}).
 \end{equation}
The proof may then be completed in the same way as the classical It\^o
theorem, making use of \ref{FEXlem}.
 \Endproof
 \section{Brownian paths and a Feynman-Kac-It\^o \break
  formula for supermanifolds}
 \label{BMSMsec}
In this section we develop a notion of Brownian paths on
supermanifolds, by extending to odd directions the constructions
of Ikeda and Watanabe \cite{IkeWat} and Elworthy
\cite{Elwort78,Elwort}. These Brownian paths on supermanifolds
will lead to a Feynman-Kac-It\^o formula for the Laplace-Beltrami
operator on forms on $M$ twisted by the bundle $E$.

Before introducing the twisting by the vector bundle, we consider
the simpler case of untwisted forms, and investigate paths on the
supermanifold $\Sm{O(M)}{T(M)}$. For the commuting, even
components of the paths, we use the construction referred to
above, that is we consider paths $x^i_t,e^i_{a,t}$ in the bundle
of orthonormal frames $O(M)$ of the manifold $M$ (with metric $g$)
starting from a point $(x^i,e^i_a)$  in $O(M)$ and satisfying the
stochastic differential equations
 \begin{eqnarray}
   x^i_t &=& x^i + \Intzt e^i_{a,s} \circ db^a_s
   \qquad e^i_{a,t} = e^i_a - \Intzt e^l_{a,s} \Chri{k}{l}{i} (x_s)
      e^{k}_{b,s} \circ db^b_s
 \label{BMSMeq}\end{eqnarray}
where $\circ$ denotes the Stratonovich product. A patching construction
allows a global solution to be constructed, since they change covariantly
under change of coordinate on $O(M)$. (Note that almost surely
 $g^{ij}(x_t)= \Sum{a=1}{m} e^i_{a\,t} e^j_{a\,t}$.)
The odd, anticommuting components of the paths
 $\xi^i_t, \pi_{i\,t}, i=1, \dots,m$
are obtained by rotating the flat fermionic paths $\theta_t, \rho_t$:
 \begin{equation}
  \xi^i_t =\xi^i + \Sum{a=1}{m} e^i_{a\,t} \theta^a_t,
 \qquad
 \pi_{i\,t}= \Sum{j=1}{n} \Sum{a=1}{m} e^j_{a\,t}g_{ij}(x_t) \rho^a_t.
 \label{ODDPATHeq}\end{equation}

In order to establish a Feynman-Kac-It\^o formula for the
Laplace-Beltrami operator $(d+\delta)^2$ on $M$, vector fields
$W_a$, $ a=1, \dots, m$, must be defined which correspond to
horizontal vector fields on $\Somu$ regarded as a bundle over $M$
with connection $\Gamma$ the Levi-Cevita connection for the metric
$g$ . In a local coordinate system $(x^i, e^i_a, \xi^i)$ on
$\Somu$ these vector fields take the form
 \begin{equation}\label{HVFUeq}
  W_a = e^i_a \Dbd{x^i} - e^j_a e_b^k \Chri{j}{k}{i}(x) \Dbd{e^i_b}
  - e^j_a \xi^k \Chri{j}{k}{i} (x) \Dbd{\xi^i} .
 \end{equation}
The key property of the vector fields $W_a$ is that, when acting
on functions on $\Somu$ which are independent of the $e^i_a$ (that
is, on functions of the form $f= g \circ \pi$ where $\pi$ is the
canonical projection from $\Somu$ onto $M$) they are related to
the Laplace-Beltrami operator $L$ by
 \begin{equation}\label{WSUeq}
  L= - \frac12 \left( W_aW_a - R_i^j(x) \xi^i \Dbd{\xi^j}
  -\frac12 R_{ki}{}^{jl}(x)\xi^i\xi^k \Dbd{\xi^j}
\Dbd{\xi^l}\right)
 \end{equation}
as may be seen from the Weitzenbock formula Lemma \ref{WFlem}
(ignoring the term involving $F$ which does not occur in the
untwisted case). This leads to the following Feynman-Kac-It\^o
formula:
 \begin{The}\label{FKUthe}
Let $(x^i_t,e^i_{a\,t}, \xi^i_t)$ be the paths defined by (\ref{BMSMeq})
and (\ref{ODDPATHeq}). Then
 \begin{eqnarray}\label{FKUeq}
 \lefteqn{ \exp(-Lt) g(x,\xi,\eta)} \End
 &=& \Exsb{
 e^{-\Intzt \Half R^j_i(x_s) \xi^i_s \pi_{j\,s}
 + \Frac14 R_{ki}{}^{jl} \xi^i_s \xi^k_s \pi_{l\,s} \pi_{j\,s} ds}
 g(x_t,\xi_t,\eta_t)}
 \end{eqnarray}
where $L=(d+ \delta)^2$ is the Laplace-Beltrami operator acting on
supersmooth functions on $\Sm{M}{TM}$.
 \end{The}
 \Oproof
Using the It\^o formula (\ref{ITOFORMULAeq}) and equation (\ref{WSUeq}),
 \begin{eqnarray}
  \lefteqn{\Exsb{g(x_t,\xi_t, \eta_t) - g(x,\xi,\eta)}} \End
&=& \Exs\bigg[\Intzt \Frac12 \left(W_a W_a
 -  R_i^j(x) \xi^i \Dbd{\xi^j}
  -\Frac12 R_{ki}{}^{jl}(x)\xi^i\xi^k \Dbd{\xi^j} \Dbd{\xi^l} \right)
 g(x_s, \xi_s,\eta_s) \bigg]ds \End
&=& \Exsb{\Intzt -L g(x_s, \xi_s,\eta_s) ds}.
 \end{eqnarray}
 Hence, if $f(x,\xi,\eta, t) = \Exsb{g(x_t, \xi_t,\eta_t)}$,
 \begin{equation}
  f(x,\xi,\eta,t) - f(x,\xi,\eta,0) =
  \Intzt -L f(x,\xi, \eta, s) \, d s
  \end{equation}
  and the result follows.
 \Endproof

To extend this Feynman-Kac-It\^o formula for the Laplace-Beltrami
operator to the twisted case, we use vector fields $\Wt_a$, $ a=1,
\dots, m$, which correspond to horizontal vector fields on $\Som$
regarded as a bundle over $M$ with connection $(\Gamma,A)$. In a
local coordinate system $(x^i, e^i_a, \xi^i, \eta^p)$ on $\Som$
these vector fields take the form
 \begin{equation}\label{HVFeq}
  \Wt_a = e^i_a \Dbd{x^i} - e^j_a e_b^k \Chri{j}{k}{i}(x) \Dbd{e^i_b}
  - e^j_a \xi^k \Chri{j}{k}{i}(x) \Dbd{\xi^i}
  -e^j_a \eta^r \Conn{j}{r}{p}(x) \Dbd{\eta^p}.
 \end{equation}
In this case we have
 \begin{eqnarray}\label{WSeq}
  L
 &=& - \frac12 \bigg( \Wt_a \Wt_a - R_i^j(x) \xi^i \Dbd{\xi^j}
  -\Frac12 R_{ki}{}^{jl}(x)\xi^i\xi^k \Dbd{\xi^j} \Dbd{\xi^l} \End
  && \quad + \Frac14 [\psi^i,\psi^j] F_{ij\,r}{}^s(x) \eta^r
  \Dbd{\eta^s}
 \bigg)
 \End
 \end{eqnarray}
as may again be seen from the Weitzenbock formula Lemma \ref{WFlem}.

In order to construct odd paths on $\Som$ we need an extended set
of commuting paths.  These are defined on the principle bundle
$O(M) \otimes E$, where $E$ is the principle bundle corresponding
to the vector bundle $E$, by the stochastic differential
equations:
  \begin{eqnarray}
   x^i_t &=& x^i + \Intzt e^i_{a,s} \circ db^a_s
   \qquad e^i_{a,t} = e^i_a - \Intzt e^l_{a,s} \Chri{k}{l}{i} (x_s)
      e^{k}_{b,s} \circ db^b_s \End
 h^p_{q\,t} &=& h^p_q +
 \Intzt h^r_{q\,s} \Conn{i}{r}{p}(x_s) e^i_{a\,s} \circ db^a_s.
 \label{BMSMTeq}\end{eqnarray}
The full set of odd paths is now obtained from fermionic Brownian paths
of dimension $m+n$, setting
 $ \xi^i_t = \xi^i + \Sum{a=1}{m} e^i_{a\,t} \theta^a_t$ and
 $ \pi_{i\,t}= \Sum{j=1}{m} \Sum{a=1}{m} e^j_{a\,t}g_{ij}(x_t) \rho^a_t$
for $i=1, \dots m$ as before, and taking
 \begin{equation}\label{TWISTEDODDPATHeq}
  \eta^p_t= \eta^p + \Sum{q=1}{n} h^p_{q\,t} \theta^{m+q}_t,
\quad \phi_{p\,t}= \Sum{q=1}{n} (h^{-1})^q_{p\,t} \rho_{m+q\,t}.
 \end{equation}
Similar arguments to those given above lead to the
Feynman-Kac-It\^o formula
 \begin{The}\label{FKthe}
 \begin{eqnarray}\label{FKeq}
 \lefteqn{ \exp(-Lt) g(x,\xi,\eta)}\hspace{-10em} \End
 &=& \Exs\Big[
 e^{\big[-\Intzt \Half R^j_i(x_s) \xi^i_s \pi_{j\,s}
 + \Frac14 R_{ki}{}^{jl} \xi^i_s \xi^k_s \pi_{l\,s} \pi_{j\,s}
 + \Frac14 \psi^i_s \psi^j_s F_{ij\,p}{}^q(x_s) \eta^p_s \phi_{q\,s}
  ds\big] } \times g(x_t,\xi_t,\eta_t)\Big] \End
 \end{eqnarray}
where $L=(d+ \delta)^2$ is the Laplace-Beltrami operator acting on
supersmooth functions on $\Sm{M}{E}$.
 \end{The}
 \section{Supersymmetric paths and the Atiyah-\hfil\break Singer
 Index Theorem}\label{ASIsec}
The Feynman-Kac-It\^o formula established in the previous section
will now be used to establish the Atiyah-Singer index theorem.
This is achieved by establishing a stronger, local version of the
theorem as in the paper of Atiyah, Bott and Patodi
\cite{AtiBotPat} for the case of the twisted Hirzebruch signature
complex.  (In \cite{AtiBotPat} it is shown by K-theoretic
arguments that this establishes the full theorem.)

The starting point is the formula of McKean and Singer
\cite{MckSin} which expresses the Hirzebruch signature of the
complex vector bundle $E$ over the Riemann manifold $(M,g)$ as the
supertrace of heat kernel of the Laplacian in the following way:
\begin{equation}\label{MSeq}
  \Ind \, (d+ \delta) = \Str \left( \exp(-Lt)\right).
\end{equation}
In the course of proving this theorem it emerges that the right
hand side is independent of $t$. The significance of this formula
in the context of supersymmetry was first appreciated by Witten
\cite{Witten82a}. Using the identification of the space of twisted
forms on $M$ with the space of supersmooth functions on $\Sm{M}{E}$, the
supertrace is defined in terms of the involution $\tau$ on the space of
supersmooth functions whose action on
$\Scap{F}$ with
 $\Scap{F}(x,\xi,\eta)= \Sum{\mu\in M_m}{} \Sum{r=1}{n}
 \Scap{F}_{\mu r}(x) \xi^{\mu}\eta^{r}$ is given by
 \begin{eqnarray}
\lefteqn{ \tau \left( \Scap{F}  \right)(x,\xi,\eta)  }\End
 &=& \Sum{\mu\in M_m}{} \Sum{r=1}{n}
 \left( \Bint d^m \rho \left( \det (g_{ij}(x)) \right)^{-\frac12}
 \exp \left( i\rho^i \xi^j g_{ij}(x)\right)
 \Scap{F}_{\mu r}(x) \rho^{\mu} \eta^r\right).
 \end{eqnarray}
This involution is essentially the Hodge dual.

The supertrace of an operator
$K$ on the space of supersmooth functions on $\Sm{M}{E}$ is then defined
(when it exists) by the formula
\begin{equation}\label{STReq}
  \Str \, {K} = \Tr\, (\tau K).
\end{equation}
The Atiyah-Singer index theorem
for the Hirzebruch signature complex can now be stated.
 \begin{The}\label{ASIthe}
 \begin{equation}\label{ASIeq}
  \Ind (d+ \delta) =
  \int_M \Mproj{\Tr \, \Exp{\frac{-F}{2\pi}}
  \det \left( \frac{i \Omega/2\pi}{\tanh i\Omega/2\pi}\right)^{\frac12}}
 \end{equation}
where as before $F$ is the curvature of the connection $A$ on the bundle
$E$ while $\Omega$ is the curvature $2$-form of the Levi-Cevita
connection on $(M,g)$, and the  brackets $\Mproj{}$ indicate projection
onto the $m$-form component of the integrand.
 \end{The}
Using the Mckean and and Singer formula \ref{MSeq}, we see that an
equivalent statement of the theorem is
 \begin{equation}\label{ASIMSeq}
 \Str (\Exp{-Lt})=
  \int_M \Mproj{ \Tr \, \Exp{\frac{-F}{2\pi}}
  \det \left( \frac{i \Omega/2\pi}{\tanh i\Omega/2\pi}\right)^{\frac12}}
\end{equation}
and thus that the theorem certainly holds if the following
stronger, local theorem holds.
 \begin{The}\label{ASILthe}
 At each point $x$ in $M$
 \begin{eqnarray}\label{ASILeq}
   \lefteqn{\lim_{t \to 0}\str (\Exp{-Lt})(x,x) d{\rm vol} }\End
 &=&
 \left.\Mproj{\Tr \, \Exp{\frac{-F}{2\pi}}
  \det \left( \frac{i \Omega/2\pi}{\tanh i\Omega/2\pi}\right)^{\frac12}
 }\right|_{x}
 \end{eqnarray}

where $\str$ denotes the $(2^mn \times 2^mn)$ matrix supertrace.
 \end{The}
The effect of the matrix supertrace is that
 $\str (\Exp{-Lt})(x,y)$ is the kernel of an operator on
smooth functions on $M$.
The proof of this theorem makes use of the Feynman-Kac-It\^o formula of
theorem
\ref{FKthe} to analyse $\exp(-Lt)$, and then (following Getzler
\cite{Getzle}), employs Duhamel's formula to extract information about the
heat kernel and show that in the limit as $t$ tends to zero only the
required term survives.

 \Proof of Theorem \ref{ASILthe}

Using (\ref{TRFeq}) we see that if $K$ is a differential operator
on the space of supersmooth functions on $\Sm{M}{E}$ then (with
$\tr$ denoting the matrix trace)
  \begin{eqnarray}
 \lefteqn{\str K(x,y) = \tr( \tau K (x,y) )  }\End
 &=&
 \Bint d^m \rho  d^m \xi d^n \eta \,
  K(x,y, \rho, -\xi,\eta,-\eta)
  \left( \det (g_{ij}(x)) \right)^{-\frac12}
 \exp \left( i\rho^i \xi^j g_{ij}(x)\right). \End
 \label{STRFCeq} \end{eqnarray}

Now the theorem is local, and it is shown by Cycon, Froese, Kirsch
and Simon in \cite{Simetal}  that at any particular point $x$ of
$M$ the manifold can be replaced by $\Real^m$ and the
$n$-dimensional vector bundle by $\Real^m \times \Comp^n$, and a
metric and connection chosen, so that in the limit as $t$ tends to
zero calculation using the Hamiltonian on $\Real^m$ gives the same
matrix supertrace as that on $M$. This is done by first choosing a
neighbourhood $W$ of $x$ which has compact closure and is both a
coordinate neighbourhood for $M$ and a local trivialization
neighbourhood for $E$. Now suppose that $U$ is also an
neighbourhood of $x$ and that $\overline{U} \subset W$. Let
$\phi:W \to \Real^m$ be a system of normal coordinates on $W$
about $x$ which satisfy $\det(g_{ij})=1$ throughout $W$. (The
existence of such coordinates are established in \cite{Simetal}.)
Additionally a local trivialization of $E$ is chosen such that
$A_{i\,r}{}^{p} (0) = 0$. (For simplicity we identify points in
$W$ by their coordinates; in particular $x$ becomes $0$.)

The standard Taylor expansion in normal coordinates about $0$ gives
\cite{AtiBotPat}
 \begin{eqnarray}
 g_{ij}(y) &=& \delta_{ij} - \frac13 y^k y^l R_{ijkl}(0)
 + \dots \End
 \Chri{i}{j}{k}(y) &=&
  \frac13 y^{l} \left( R_{lji}{}^{k}(0) + R_{lij}{}^{k}(0)\right)
  + \dots \End
  A_{i \, r}{}^{p} (y) &=& - \frac12 y^j F_{ij \, r}{}^{p} + \dots.
 \label{NCeq}\end{eqnarray}
Now let $\Hf$ be the flat Laplacian $\Dbd{x^i}\Dbd{x^i}$ on
$\Real^m$ and let $K_t(y,y',\xi,\xi',\eta,\eta')$ and
$K^0_t(y,y',\xi,\xi',\eta,\eta')$ be the heat kernels of $L$ and
$\Hf$ respectively.  From Duhamel's formula (as quoted by Getzler
in \cite{Getzle}) we learn that
 \begin{eqnarray}
 \lefteqn{ K_t(y,y',\xi,\xi',\eta,\eta') -
 K^0_t(y,y',\xi,\xi',\eta,\eta')}\End
 &=&\Intzt ds\, e^{-(t-s)L} (L-\Hf) K^0_s(y,y',\xi,\xi',\eta,\eta')
 \end{eqnarray}
where the differential operators $L$ and $\Hf$ act with respect to the
unprimed arguments. Now
 \begin{eqnarray}
 \lefteqn{  K^0_s(y,y',\xi,\xi',\eta,\eta')}\End
 &=&\Bint d^m{\rho} d^n{\kappa} (2\pi s)^{\frac{m}{2}}
 \Exp{-(y-y')^2/2s -i\rho(\xi-\xi')-i\kappa(\eta-\eta')}.
  \end{eqnarray}
From this we may deduce by direct calculation that $\str K^0_t(0,0) = 0$
and, using (\ref{STRFeq}),
 \begin{eqnarray}
 \str K_t(0,0) = \qquad && \End
 \Bint d^m \xi d^m \xi' d^n \eta
 \,\exp{-i\xi \xi'}
 & \times & \Intzt ds \bigg[ \left( e^{-(t-s)L}(L-\Hf)
 K^0_s(0,0,\xi,\xi', \eta, \eta)\right)
   \bigg]. \End
 \end{eqnarray}
The Feynman-Kac-It\^o formula theorem \ref{FKthe} can then be used
to evaluate this expression.  Rather than proceeding directly, we
now construct a simplified Hamiltonian $\Hs$ on $\Real^m \times
\Comp^n$ with heat kernel $K^1_t$ which has the property
 \[
  \lim_{t \to 0} \str K^1_t(0,0) =  \lim_{t \to 0} \str K_t(0,0)
 \]
so that the required supertrace can be calculated.

The modified Laplacian is built from the approximate metric and
connections of (\ref{NCeq}), working to first order in $y$, and with
Euclidean metric. Writing $R_{ijkl}{}^{l}$ for
$R_{ijkl}{}^{l}(0)$ and $F_{ij\,p}{}^{q}$ for $F_{ij\,p}{}^{q}(0)$
it is
 \begin{eqnarray}
 \lefteqn{ \Hs = -\bigg[ \frac12 \Dbd{x^i} \Dbd{x^i}
 -\frac14 R_{ij}{}^{kl} \theta^i\theta^j \Dbd{\theta^l} \Dbd{\theta^k}
 + \frac13 R^k_j \theta^j \Dbd{\theta^k}  } \End
 && + \frac14 \psi^i \psi^j \eta^p F_{ij\,p}{}^{q} \Dbd{\eta^q}
  -\frac13 \theta^jx^l\left( R_l{}^k{}_j{}^i + R_{lj}^{ki}\right)
  \Dbd{\theta^i} \Dbd{x^k} \End
  &&- \frac1{18} \theta^{k'}\theta^{l'} x^k x^l
   \left( R_{k j' k'}{}^i +  R_{k k' j'}{}^i \right)
   \left( R_l{}^{j'}{}_{l'}{}^j + R_{l l'}{}^{j' j} \right)
   \Dbd{\theta^i} \Dbd{\theta^j} \bigg].
 \end{eqnarray}

The brownian paths which lead to a Feynman-Kac-It\^o formula as in
\ref{FKthe} are
 \begin{equation}
 x^1{}^i_t = b^i_t, \qquad
 \xi^1{}^{i}_t = \xi^i + \theta^i_t, \qquad
 \eta^1{}^p_s = \eta^p +  \theta^{m+p}_t.
  \end{equation}
Using these paths in the Feynman-Kac-It\^o formula, it can then be shown
(the details are in \cite{SCSTWO}) that
 \begin{equation}
 lim_{t \to 0} K_t(0,0) =lim_{t \to 0} K^1_t(0,0).
 \end{equation}

By evaluating the supertrace of $\exp(-\Hs t)$ (using flat space path
integration techniques for both fermionic and bosonic paths)the local
form of the Atiyah-Singer index theorem \ref{ASILthe} will be
established.

Using Duhamel's formula and changing variable
 $\phi = \sqrt{2\pi t}\,\xi$ gives
 \begin{eqnarray}
 \lefteqn{K^1_t(0,0)} \End
  &=& \Exs\Bigg[\Bint d^m\phi d^n \eta d^n\kappa
  d^m \xi  (2\pi t)^{\frac{m}{2}}
   \, \Intzt ds (2\pi s)^{-\frac{m}{2}}  \End
  && \times \Big\{\Exp{-\Hs(t-s)}(\Hs -\Hf)
  \Scap{G}_s(0, \frac{\phi}{\sqrt{2 \pi t}},\xi',\eta,-\eta) \End
  &&\times \Exp{-i\kappa\eta-i\frac{\phi}{\sqrt{2 \pi t}}\xi'}
   \Big\}
    \Bigg]\End
  \end{eqnarray}
  where
   \begin{equation}
  \Scap{G}_s(x,\xi,\xi',\eta,\kappa)
  = \exp-(x^2/2s +i\xi.\xi' + i\kappa.\eta)
 \end{equation}
 so that
\begin{equation}
 \Bint d^n\kappa  \Scap{G}_s(x,\xi,\xi',\eta,\kappa) =
   \tau K^0_s(x,0,\xi,\xi',\eta,0).
 \end{equation}
Hence, using the fact that (for small $t$) $x_t \sim \surd t$ while the
fermionic paths are of order $1$ \cite{SCSTWO} it can be shown that
  \begin{eqnarray}
 \lefteqn{\lim_{t \to 0}K^1_t(0,0)} \End
  &=& \Exs\Bigg[\Bint d^m\phi d^n \eta d^n\kappa d^m \xi
  (2\pi t)^{\frac{m}{2}}
   \, \Intzt ds (2\pi s)^{-\frac{m}{2}} \End
 && \times \Big\{ \left( \Ho-\Hf \right)
 G_s(b^i_{t-s}, \theta^i_{t-s},\frac{\phi^i}{\surd{2 \pi t}},
 \theta^{m+p}_{t-s},\kappa^p)
  \exp i\kappa .\eta \Big\}\Bigg]
   \end{eqnarray}
where $\Ho=\Ho{}_{x} + \Ho{}_{\xi} +\Ho{}_{\eta}$ with
 \begin{eqnarray}
  \Ho{}_{x} &=& -\Big( \frac12 \Dbd{x^i}\Dbd{x^i}
  + \frac12 x^l \frac{\phi^j\phi^i}{{2 \pi t}}
  R_{jil}{}^k \Dbd{x^k}
  + \frac18 x^k x^l \frac{\phi^i\phi^j\phi^{i'}\phi^{j'}}{(2\pi t)^2}
  R_{i'ikk'}R_{j'jl}{}^{k'} \Big)
  \End
   \Ho{}_{\xi}&=& -\frac14  R_{ijkl} \frac{\phi^j\phi^i}{{2 \pi t}}
   \psi^{k}\psi^{l}
  \End
   \Ho{}_{\eta}&=& - \frac{\phi^j\phi^i}{{2 \pi t}}
   F_{ij\,p}{}^{q} \eta^p \Dbd{\eta^q}.
   \end{eqnarray}
Now $\Exp{-\Ho{}_x t(0,0)}$ can be evaluated using the standard
 $\Real^2$ result \cite{Simon}  that if
 \[
 L= - \frac12\Dbd{x^i}\Dbd{x^i}
 + \frac{iB}{2}\left( x^1 \Dbd{x^2} - x^2 \Dbd{x^1}\right)
 + \frac18 B^2 \left( (x^1)^2 + (x^2)^2\right)
 \]
then
 \begin{equation}
  \Exp{-Lt(0,0)}= \frac{B}{4\pi \sinh(\frac12Bt)}.
 \end{equation}
Skew-diagonalising the
 $\Omega_k{}^l = \frac12 \phi^i\phi^j R_{ijk}{}^l$ by $2$ by $2$ blocks
 $\Matt{0}{\Omega_k}{-\Omega_k}{0},k=1, \dots \frac12 m$
down the leading diagonal we see that
 \begin{equation}
  \Exp{-\Ho{}_x t(0,0)}= \prod_{k=1}^{m/2}
  \frac{i \Omega_k}{2 \pi t} \,
   \frac{1}{\sinh(i \Omega_k/2\pi)}.
 \end{equation}
Also, using flat fermionic paths or direct calculation we see that
 \begin{eqnarray}
 \lefteqn{\Exp{-\Ho{}_{\xi} t(\xi,\xi')}=
  \Bint d^m \rho \Big\{
 \Exp{-i\rho(\xi-\xi')} } \End
  && \times \prod_{k=1}^{m/2}
 \big( \cosh \frac{i\Omega_k}{2\pi}
 + (\xi^{2k-1}+i\rho^{2k-1})(\xi^{2k}+i\rho^{2k})
 \sinh \frac{i\Omega_k}{2\pi} \big)  \Big \}.
 \end{eqnarray}

This leads to
 \begin{eqnarray}
 \lefteqn{\str\Exp{-\Ho{} t(0,0)}=} \End
 &=& \Bint d^m \phi \prod_{k=1}^{m/2}
  \left(\frac{i \Omega_k}{2 \pi t}
   \frac{1}{\sinh(i \Omega_k/2\pi)}
   \cosh \frac{i\Omega_k}{2\pi} \right)
   \tr \left( \exp \left( -\phi^i \phi^j \frac{F_{ij}}{2\pi}\right)\right).
 \end{eqnarray}
so that
\begin{equation}
   \str\Exp{-L t(0,0)}\, d{\rm vol}=\Mproj{
  \tr \left( \exp \left( - \frac{F}{2\pi}\right)\right)
  \det\left(\frac{i \Omega/2 \pi }{\tanh(i \Omega/2\pi)}
  \right)^{\frac12}}.
 \end{equation}
 as required.
 \Endproof
 \section{Superpaths and the index of the Dirac operator}\label{DIRACsec}
In this section we introduce super Brownian paths on the spin
supermanifold $\Spinsm$, parametrised by a commuting parameter $t$ and an
anticommuting parameter $\tau$.  An  It\^o theorem is established for
these paths, which is applied to give a path integral formula for the
kernel of the super evolution operator
 $\Exp{\Dira^2 t + \Dira\tau}$.

Suppose that $p$ is a point on the spin supermanifold $\Spinsm$
with coordinates $(x;\xi)$.  The superpath based at $p$ is
constructed from solutions to the following system of purely
bosonic stochastic differential equations
 \begin{eqnarray}
 x^i_t &=& x^i + \Intzt \Mm{b}{a}{s} e^i_b(x_s) \circ db^a_s,
 \quad i=1, \dots,m \End
 \Mm{b}{a}{t} &=& \delta^b_a +
    \Intzt \Mm{d}{e}{s} \Mm{c}{a}{s} \Chri{d}{c}{b}(x_s) \circ db^e_s
    - \Intzt \Frac12 \Mm{c}{a}{s} R^b_c \, ds,
    \quad a,b = 1, \dots , m. \End
 \end{eqnarray}
The superpath $X\Subtt=(x\Subtt;\xi\Subtt)$ is then defined by
 \begin{eqnarray}
 x^i\Subtt &=& x^i_t + \Frac{i}{\surd2} \tau \xi^a_t e^i_a(x_t),
 \qquad i=1, \dots, m \End
 \xi^a\Subtt &=& \xi^a_t
 + \Frac{i}{2\surd2} \tau \Gamma^{a}{}_{bc} \xi^b_t \xi^c_t,
 \qquad a=1, \dots,m
 \End
 \end{eqnarray}
where $\xi^a_t = \omega^b_t \Mm{a}{b}{t}$.
The It\^o formula for these paths, generalising the super integration
formula (\ref{INTALONGSUPERPATHeq}), is the content of the following
theorem:
 \begin{The}
 Suppose that $F$ is supersmooth on $\Spinsm$. Then
\begin{equation}\label{SUPERITOeq}
  f(X\Subtt)-f(X_0)
  \Muq
  \Intzztt Dx^i \Ds{i} f(X\Subss) +D \xi^a \Ds{m+a}f(X\Subss)
\end{equation}
where
 \begin{eqnarray}
 DX^i &=& \Dss \, \Frac{i}{\surd2} e^i_a(x_s) \xi^a_{s^-}, \End
 D\xi^a &=& \Dss \, \Frac{i}{2\surd2}
 \Gamma^a{}_{bc} \xi^b_{s^-} \xi^c_{s^-}
 +d\sigma \, \sigma d\xi^a_{s^-}.
 \end{eqnarray}
(Here the symbol $s^-$ indicates that the path $\xi^a_s$ is to be
taken at a time infinitesimally before $s$.)
 \end{The}
This theorem may be proved by expanding in powers of $\tau$.  It
leads immediately to the supersymmetric Feynman-Kac-It\^o formula
for the Dirac operator.
 \begin{The}
\begin{equation}\label{SUPERFKeq}
  \Exp{-\Dira^2 t -\Dira \tau} f(p)
  = \Exsb{f(X\Subtt)}.
\end{equation}
 \end{The}
 \Oproof  Expanding the right hand side of (\ref{SUPERITOeq}) gives
\begin{eqnarray}
 \lefteqn{    \Exsb{\Intzztt Dx^i \Ds{i} f(X\Subss)
 +D \xi^a \Ds{m+a}f(X\Subss)} } \End
 &=&
 \Exsb{ \Intzztt \Dss \Dira f(X\Subss)
 -2 \sigma \Dira^2f(X\Subss)}.
\end{eqnarray}
If we now define the operator $U\Subtt$ by
\begin{equation}
  U\Subtt f(p) = (Pg)(p) \quad \mbox{where} \quad
  g(p) = \Exsb{f(X\Subtt)}
\end{equation}
we find that
 \begin{equation}
  U\Subtt f - U_{0;0} f
  = \Intzztt \Dss U\Subss(\Dira f - 2 \sigma \Dira^2 f)
 \end{equation}
so that
 \begin{equation}
  D\Subtt U\Subtt f(x;\xi) = U\Subtt (\Dira-2\tau \Dira^2) f(x;\xi).
 \end{equation}
Using the fact that
 \begin{equation}
  D\Subtt \left(\Exp{-\Dira^2 t - \Dira \tau} \right)
  = \Exp{-\Dira^2 t - \Dira \tau} (\Dira - 2 \tau \Dira^2),
 \end{equation}
we see that $U\Subtt = \Exp{-\Dira^2 t - \Dira \tau}$ as required.
\Endproof
  \section{Ghosts and the quantization of systems\hfil\break with symmetry}\label{BRSTsec}
When using functional integrals to quantize a system with gauge
symmetries, some mechanism must be found to remove the redundancy
which arises from gauge equivalence.  A widely-used method to
achieve this uses a Lagrangian modified by gauge-fixing and
Faddeev-Popov ghost terms. The canonical quantization of such
theories was described by Batalin, Fradkin, Fradkina and
Vilkovisky in a seies of papers
\cite{FraVil1,BatVil,FraFra,BatFra,FraVil2}.  (The methods
developed in these papers will be refereed to as the \textsc{BFV}
approach.) An elegant interpretation of this work, in particular
an explanation of the rather complex gauge-fixing terms, was given
by Henneaux in terms of BRST cohomology in \cite{Hennea}. A key
idea in these methods is that the original Hamiltonian of the
theory must be adjusted by a term which is the anticommutator of
two operators (both odd), the BRST charge $\Brst$ of the theory
(which determines the BRST cohomology of the theory) and the
gauge-fixing fermion $\Gff$.  The r\^ole of the gauge-fixing
fermion in ensuring the validity of the gauge-fixing procedure in
the path integral was clarified by the author in \cite{GFBFVQ}.

In the canonical approach symmetries of a system appear as
constraints on the canonical positions and momenta. The simplest
situation is that where the constraints are `first class', that
is, in involution under the Poisson bracket of the system and when
acting on the Hamiltonian.  For definiteness suppose that we have
an unconstrained $2n$-dimensional phase space $\Real^{2n}$, (with
local coordinates $(p_i,q^i)$, where $i=1,\dots, n$) together with
a set of $m$ first class constraints $T_a(p,q)=0, a=1,\dots,m$
(with $m<n$) and first class Hamiltonian $H_c{}(p,q)$.

The true (reduced) phase space of this system is then the space $B=C/G$
where $C$ is the submanifold of $\Real^{2n}$ on which the constraints
hold and $G$ denotes the group generated by the constraints (which acts
naturally on $C$), with symplectic structure given by the Dirac bracket.
A set of gauge-fixing functions $X^a, a=1,\dots,m$ are also introduced,
whose essential properties are described below.

Now, as is shown in \cite{Faddee}, the generating functional of the
theory is given (in the more informal language of path integrals used in
physics) by the Faddeev formula
 \begin{eqnarray}
   Z &=& \int \mathcal{D} p   \mathcal{D} q   \Bigg[ \prod_t
 \left(\prod_{a=1}^m \left[ \delta \left(T_a(p(t),q(t)) \right) \delta \left (X^a(p(t),q(t)) \right)  \right] \right)\End
 && \times \det \left(\{ T_a(p(t),q(t)),X^b(p(t),q(t)) \}\right) \End
 && \times \exp \left(  i\int_0^t p_a(t) \dot{q}^a(t)   - H_c{}(p(t)  ,q(t)) dt  \right)\Bigg],
 \label{FFeq}\end{eqnarray}
where the integration  is over paths $p(t), q(t)$ in the
unconstrained phase space which begin and end at the same point.
(Standard quantization of the unconstrained phase space in the
Schr\"odinger picture is used.) In establishing this expression,
further assumptions have to be made about the gauge-fixing
functions $X^a$, including the requirement that the matrix
$(\Pb{T_a}{X^b} )$ (where $\Pb{ \hspace{7pt}}{ \hspace{7pt}}$ denotes
the Poisson bracket) must satisfy the condition
 \begin{equation}
  \det (\{ T_a,X^b \} ) \not= 0
  \label{PBeq}
 \end{equation}
at all points $(p,q)$  in $\Real^{2n}$. (In the approach described
below alternative, wider, possibilities for gauge-fixing are
described \cite{GFBFVQ}.)

In the BFV  approach the phase space is extended by including Lagrange
multipliers $l^a, a=1,\dots,m$ for the constraints together with their
canonically conjugate momenta $k_a$, a set of $m$ ghosts, $\eta^a$
together with their conjugate momenta $\pi_a$ and a set of $m$ antighosts
and corresponding momenta $\phi_a$ and $\theta^a$. An extended
Hamiltonian $\Hame = H_c{}  +\Hamg$ is then defined, with $\Hamg$ taking
a form prescribed by the constraints, their commutators, and the
gauge-fixing functions. The extended Hamiltonian is quite complicated,
but it is shown in \cite{FraVil2}, by clever manipulation of path
integrals, both that the corresponding generating functional
 \begin{eqnarray}\label{FVSSeq}
  Z &=& \int \mathcal{D} p   \mathcal{D} q   \mathcal{D} k   \mathcal{D} l   \mathcal{D}
    \pi   \mathcal{D} \eta   \mathcal{D} \phi   \mathcal{D} \theta   \End
  &&  \times \Bigg[ \exp \bigg(i \int_0^t p(t)  \dot{q} (t)  + k (t)
    \dot{l}(t) +  \pi(t) \dot{\eta}(t)   +   {\phi}(t)    \dot{\theta}(t)   \quad  \End
    &&\qquad \qquad  -  \Hame(p(t)  ,q (t) ,k (t) , l  (t), \pi (t) ,
   \eta(t), \phi(t) , \theta(t)  ) dt\bigg) \Bigg] \End
  &&
 \end{eqnarray}
is independent of the choice of gauge-fixing functions and that it
is equal to the Faddeev formula (\ref{FFeq}) for the generating
functional for the original Hamiltonian on the reduced phase
space. This result is known as the {\em Fradkin-Vilkovisky
Theorem}.  A significant result given by Henneaux \cite{Hennea}
was an interpretation of the BFV Hamiltonian $\Hame = H_c{} +
 \Hamg$ in terms of the cohomology of the BRST  operator
$\Brst$ corresponding to the $2(m+n)$-{}dimensional phase space
(with typical point $(p,q,l,k)$) subject to the constraints
$T_a(p,q)=0,\, k_a=0$. Henneaux showed that  the term $\Hamg$
could be expressed as $i\Comm$ with a gauge-fixing fermion $\Gff$
constructed from the gauge-fixing functions $X^a, a= 1, \dots, m$
in a prescribed way. Henneaux also demonstrated the correspondence
between observables on the reduced phase space and operators which
commute with $\Brst$, and the related correspondence between
states for the reduced system and $\Brst$-cohomology classes.

In \cite{GFBFVQ} it is shown that the role of the anticommutator
$i\Comm$ is to regulate the generating functional of the theory,
ensuring that the summation involved is absolutely convergent so
that the necessary cancellations which occur, so that the trace
projects onto physical states only. This will occur if $\Gff$ is
chosen so that on each ghost and $\Hame$ sector  the eigenvalues
of $\Comm$ tend to infinity and $\Hame$ has a finite trace on the
space of zeroes of $\Comm$ at each ghost number. The key steps in
the argument are first to recall the argument of Schwarz
\cite{Schwar} showing that the supertrace of an observable on the
space of states of the theory is equal to the alternating sum of the
traces over BRST cohomology classes, and then to observe that the path
integral will give a supertrace. An example of this mechanism is
described in the following section. The method is also valid when the
phase space is extended merely by including ghosts $\eta^a$ and their
conjugate momenta $\pi_a$, provided that a gauge-fixing fermion can be
found with the necessary properties.
 \section{Morse theory and the topological particle}\label{MTTsec}
This section concerns the quantization of a highly symmetric
quantum mechanical system, the topological particle.  Starting
from the classical action, which involves a single function $h$ on
a manifold $M$, the canonical constraints are derived and the BRST
Hamiltonian constructed. This Hamiltonian is the supersymmetric
Hamiltonian given by Witten in is analysis of Morse theory
\cite{Witten82}. The path integral methods developed earlier in
this article are used to calculate the heat kernel (or matrix
elements of the evolution operator) between critical points of
$h$, which are central to Witten's arguments. The key technique is
to define paths by a stochastic differential equation which encodes
fluctuations about classical trajectories and thus gives a rigorous
version of the phase space WKB method (described by Blau, Keski-Vakkuri
and  Niemi in \cite{BlaKesNie90}).

The topological particle is a model which was first  described by
Baulieu and Singer \cite{BauSin}. The field of the theory is a
smooth map $x: [0,t] \to M$ (where as before $M$ is a smooth,
compact manifold of dimension $m$), and the action is
  \begin{equation}\label{ACeq}
  S\left(x(.)\right) = \Intzt \sum_{{j}=1}^m v_{j}(x(t'))\, \dot{x}^{j}(t') \, d t'.
 \end{equation}
 where $v=dh$ is an exact $1$-form with $h: M \to \Real$ a Morse
 function.  The components $v_{j}$ relative to local coordinates
  $x^{j}$ on $M$ are as usual defined by
  $v = \sum_{{j}=1}^m v_{j}  dx^{j}$ so that
  $v_{j} = \Dbdf{x^{j}}{h}$.  This action can of course be
expressed in the much simpler form
$ S\left(x(.)\right) = h(x(t))- h(x(0)) $, and this form makes it
clear that the action is invariant under arbitrary changes of the field
$x$ provided that the endpoints remain fixed.  This is the first
indication that the theory is a so-called topological theory, a theory
with no independent degrees of freedom and whose equations of motion are
identically satisfied.  One might assume such a theory was of no
interest, but in fact, although at the linearized level there is no
dynamics, on quantization topological information is captured.

The first step in quantization is to investigate the Hamiltonian
dynamics of the model.  Using the action (\ref{ACeq}) we see that
the Lagrangian  is
 \begin{equation}\label{LagrangianEq}
  \Lag (x,\dot{x}) = v_{j}(x) \dot{x^{j}}
 \end{equation}
 so that, using the Legendre transformation to the phase space
 $T^*(M)$ the canonical momentum is
 \begin{equation}
  p_{j}= i \frac{\delta \Lag(x,\dot{x})}{\delta \dot{x}^{j}}
  = i v_{j}(x)
 \end{equation}
where we are using Euclidean time, which differs from the usual
physical time of quantum mechanics by a factor $i$.

The symmetries of the system now reveal themselves as
$m$ constraints on the canonical position and momenta $x^{j}, p_{j}$
of the form
 \begin{equation}\label{CONSTRAINTSeq}
  T_{j} \equiv -p_{j}+ i v_{j}(x) = 0, \qquad {j} = 1, \dots, m.
 \end{equation}
The Poisson brackets on the phase space $T*(M)$ are obtained from
the standard symplectic form $\omega = dp_{i} \wedge dx^{j}$,
leading to
 \begin{equation}\label{FIRSTCLASSeq}
  \Pb{T_{j}}{T_{k}} = 0, \qquad j,k = 1, \dots, m,
 \end{equation}
where we have used the fact that $v$ is closed.

The standard prescription determines the Hamiltonian to be
 \begin{equation}\label{HAMCeq}
  H = ip_{j} x^{j} + \Lag(x,\dot{x}) = 0
 \end{equation}
so that the constraints are first class  and abelian. The number of
constraints equals the number of fields, which is a further indication of
the topological nature of the theory; by simple counting one would expect
all field configurations to be equivalent.  The first hint that something
more survives comes from the gauge-fixing process; a natural choice of
gauge-fixing conditions is
 \begin{equation}\label{GAUGEFIXINGeq}
  X^{j}= g^{{i}{j}}(-p_{ij}- i v_{i}),
 \end{equation}
where $g$ is a Riemannian metric on $M$.  (Justification for this choice,
and the  independence of its consequences on the particular Riemannian
metric used, will be made when the model is quantized below.)

To implement the constraints and gauge fixing in the quantum theory we
use the BRST approach, introducing ghosts $\eta^{j}$ and their canonical
momenta $\pi_{j}, j=1, \dots m$. The configuration space for the BRST
theory is the supermanifold $\Sm{M}{T(M)}$ with coordinates $x^{j},
\eta^{j}$ corresponding to the original fields of the theory and the
ghosts. The phase space is $\Sm{T^*(M)}{T(M) \times T^*(M)}$, where
$T(M)$ and $T^*(M)$ are regarded as bundles over $T^*(M)$ by pulling back
from $M$ using the projection map $T^*(M) \to M$. The even coordinates on
phase space are $x^{j}, p_{j}$ while the odd coordinates are $\eta^{j},
\pi_{j}$ which correspond to the ghosts and antighosts respectively.
The `classical' BRST fields are maps from $[0,t]$ into the configuration
supermanifold.  The simplest choice of symplectic form on the super phase
space is
 \begin{equation}\label{SUPERSYMPLECTICeq}
  \omega_{s} = dp_{j} \wedge dx^{j} + D\pi_{j} \wedge D \eta^{j}
  - \Half R_{{i}{j}{k}}{}^{l} \pi_{l} \eta^{k} dx^{i} \wedge dx^{j}
 \end{equation}
where $R$ is the Riemann curvature of $g$ and $D$ denotes covariant
differentiation using the Levi-Civita connection so that
 \begin{equation}\label{DETAeq}
  D\eta^{j} = d \eta^{j} + \Chri{i}{k}{j} \eta^{k} dx^{i}
  \qquad \Mbox{and} \qquad
  D\pi_{k} = d \pi_{k}  - \Chri{i}{k}{j} \pi_{j} dx^{i}.
 \end{equation}
The corresponding Poisson brackets are
 \begin{eqnarray}\label{SPBeq}
   \Pb{p_{j}}{x^{i}} &= \delta^i_j
   \qquad \Pb{p_{i}}{p_{j}} &= R_{{i}{j}{k}}{}^{l} \, \pi_{l} \eta^{k} \End
   \Pb{p_{j}}{\eta^{i}} &= \Chri{j}{k}{i} \eta^{k}
   \qquad  \Pb{p_{j}}{\pi_{i}} &= -\Chri{j}{i}{k} \pi_{k} \End
   \Mbox{and} \quad & \Pb{\pi_{j}}{\eta^{i}} &= \delta^i_j,
 \end{eqnarray}
with others being zero.

To quantize this system in the Schr\"odinger picture we use states which
are supersmooth functions on $\Sm{M}{T(M)}$, or, in more physical
language, wave functions $\psi(x,\eta)$.  The position observables $x$
and $\eta$ are represented by multiplication while the momentum
observables are represented by covariant differentiation:
 \begin{equation}\label{SREPeq}
  p_{j} = -i D_{j} \equiv -i \left(\Dbd{x^{j}} +
  \eta^i \Chri{i}{j}{k} \Dbd{ \eta^{k}}\right) \quad \Mbox{and} \quad
  \pi_{j} = -i  \Dbd{\eta^{j}}.
 \end{equation}
The BRST operator is (following the standard procedure)
 \begin{equation}\label{BRSTeq}
  \Brst = \eta^{j} T_{j}
  = i \eta^{j} \left( \Dbd{\eta^{j}} + v_{j}(x) \right),
 \end{equation}
while the gauge-fixing fermion is
 \begin{equation}\label{GFFeq}
  \Gff = \pi_{j} X^{j}
  = i g^{{j}{k}} \pi_{j} \left( D_{k}- v_{k}(x)  \right).
 \end{equation}
Identifying functions on configuration space with forms on $M$, we see
that
 \begin{equation}
    \Brst =   i e^{-h} d e^{h}
    \qquad \Mbox{and} \qquad
    \Gff = e^{h}\delta e^{-h}
 \end{equation}
where $d$ is exterior differentiation and $\delta = * d *$ is the adjoint
of $d$. Thus we see that $\Brst$ and $\Gff$ are the supersymmetry
operators introduced by Witten in his study of supersymmetry and Morse
theory \cite{Witten82}.

The BRST Hamiltonian for the theory is simply $\Com{\Brst}{\Gff}$, since
the classical Hamiltonian is zero.  It is thus equal to the Hamiltonian
used by Witten in \cite{Witten82}, and has the explicit form
 \begin{equation}\label{WHAMeq}
  H = \Half ( d + \delta )^2
  + \Half g^{{j}{k}} \Dbdf{h}{x^{j}} \Dbdf{h}{x^{k}}
  + \Half g^{{j}{k}} (\eta^{l} \pi_{j} - \pi_{j} \eta^{l})
  \frac{D^2 h}{Dx^{k} Dx^{l}}.
 \end{equation}
Using the mapping $\psi \mapsto e^{-h} \psi$, Witten shows that
the cohomology of $\Brst$ is isomorphic to the De Rham cohomology
of $M$, and that forms with zero $H$ eigenvalues give exactly one
representative of each $\Brst$ cohomology class.  Thus the
gauge-fixing used is a good one, and leads to results which are
independent of the choice of metric.

To describe how the superspace path integral techniques developed
in earlier sections can be applied to this section, we begin for
simplicity by working in flat space, with the Euclidean metric.
Even in this situation the key new idea can be illustrated; this
is to use a stochastic differential equation which leads to paths
which encode directly Brownian fluctuations about the classical
trajectories, giving a
stochastic version of the WKB method. 
In the standard WKB approach only second order fluctuations are
considered, first order fluctuations vanishing because they are
taken about classical trajectories while higher order fluctuations
are ignored. The stochastic formula we present here is exact.

Consider first the stochastic differential equation
 \begin{equation}\label{NMeq}
  dx^{j}_t = db^{j}_t - v_{j}(x_t) dt, \quad x_0 =x
 \end{equation}
which is the equation for the classical trajectories augmented by
stochastic fluctuations; this corresponds to the Nicolai map of the
theory.  For positive $t$ we consider the operator $U_t$ acting on
functions $\psi$ on configuration space given by
 \begin{equation}\label{UTeq}
  U_t \psi (x ,\eta) =
  \Exsb{e^{ \Intzt (\Ds{j} h(x_s) dx^{j}_{s} + \Ds{j} \Ds{k} h(x_s)
  \left( i \theta_{s}^{j} \rho_{s \, {k}} + \Half \delta^{j}_{k} \right)
  ds)}
  \psi(x_t, \eta_t)  }.
 \end{equation}
By It\^o calculus we see that
 \begin{eqnarray}
  \lefteqn{ \left[ \Exp{ \Intzt (\Ds{j} h(x_s) dx^{j}_{s} + \Ds{j} \Ds{k} h(x_s)
  \left( i \theta_{s}^{j} \rho_{s \, {k}} + \Half \delta^{j}_{k} \right)
  ds )
  \psi(x_t, \eta_t)  } \right] } \End
  &=& \left[ e^{ \Intzt (\Ds{j} h(x_s) dx^{j}_{s} + \Ds{j} \Ds{k} h(x_s)
  \left( i \theta_{s}^{j} \rho_{s \, {k}} + \Half \delta^{j}_{k} \right)
  ds )}
 (-H) \psi(x_t, \eta_t)   \right] \End
 && + \quad  \Mbox{terms of zero measure},
 \end{eqnarray}
so that
 \begin{equation}
  \Dbdf{U_t\psi(x, \eta)}{t} = - U_t H \psi(x,\eta)
 \end{equation}
and so we may conclude that
 \begin{equation}
  U_t = \exp -tH.
 \end{equation}
Making further use of It\^o calculus we can simplify the expression for
$U_t$: we have
 \begin{equation}
  \Intzt \left( \Ds{j}h (x_s) dx^{j}_s +
  \Half \Ds{j} \Ds{j} h(x_s) ds \right)
  = h(x_s) - h(x)
 \end{equation}
so that equation (\ref{UTeq}) can be rewritten
 \begin{eqnarray}\label{FKHeq}
  \lefteqn{ \exp -t H \psi(x,\eta)}  \End
  &=& \Exsb{ \Exp{-(h(x)-h(x_t))}
  \Exp{\Intzt i\Ds{j} \Ds{k} h(x_s) \theta^{j}_s \rho_{s\,{k}} ds}
  \psi(x_t, \theta_t)  }. \End
 \end{eqnarray}
This expression shows how the WKB factor $\exp -\Delta h$ (which
corresponds directly to the original topological action (\ref{ACeq}))
appears in the path integral.

This formula is readily adapted to curved space: in place of \ref{NMeq}
one simply uses the covariant stochastic differential equation
 \begin{equation}\label{NMCeq}
  d \Xt^{j}_t= dx^{j}_t - g^{{j}{k}} (\Xt_t) v_{k}(\Xt) dt
 \end{equation}
with $x_t$ the curved space Brownian paths of (\ref{BMSMeq}), and then
obtains the Feynman-Kac-It\^o formula
    \begin{eqnarray}\label{FKIeq}
 \lefteqn{\exp-t\Hamg \psi(x,\eta) = \int d\mu \exp( - (h(x)-h(  x_t))) } \End
 &&
 \exp \Big(\Intzt \big( {D_{j}D_{ k}h({  x_s})}
  i   g^{l k}(  x^{j}_s)\theta^{j}_s \rho_{s\,l}
   \End
 &+& R_{j}^{k}(  x_s) \theta^{j}_s \rho_{k\, s}
 +\Frac12 R_{j k}^{l i}(  x_s)
 \theta^{j}_s \theta^{k}_s \rho_{l\, s}\rho_{i \, s}
          \big)ds \Big) \psi(  x_t,\theta_t).
   \end{eqnarray}

much as before.

In \cite{Witten82} Witten rescales the Morse function $h$ by a factor
(here denoted by $u$) and by taking the large $u$ limit distils out
information about the topology of the manifold, explicitly building the
cohomology of the manifold in terms of the critical points of $h$.  The
stochastic methods developed here allow a rigorous mathematical version
of Witten's calculations to be derived in the spirit of the physicist's
approach.  Some of Witten's work is derived in a mathematically rigorous
way by Cycon, Froese  Kirsch and Simon in \cite{Simetal}, we make use of
their results here.  A full mathematical derivation of Witten's result is
given by Helffer and Sj\"orstrand in \cite{HelSjo85}, using semiclassical
techniques.

It is useful at this stage to introduce some standard terminology
and results from Morse theory: a critical point $a$ of the
function $h$ is said to have index $p$ if the Hessian matrix
$\left( D_{x^{j}} D_{x^{k} h} \right)$
 has exactly $p$ negative eigenvalues. Each critical point $a$ has a
neighbourhood $N_a$ on which a special coordinate system can be chosen in
which the Morse function $h$ takes the standard form
 \begin{equation}\label{MORSECOORDeq}
  h(x) = h(a) +
  \Half \sum_{{j}=1}^m \sigma_{j} (x^{j} - a^{j})^2
 \end{equation}
where $\sigma_{j} = +1$ for ${j}=1, \dots, m-p$ and $\sigma_{j} =
- 1$ for ${j}=m-p+1, \dots, m$.  A metric $g$ on $M$ satisfies the
Smayle transversality condition for $h$ if the solution curves
$\Gamma_{ab}$ to the `steepest descent' differential equation
 \begin{equation}\label{STeq}
 \frac{dx^{\mu}(t)}{dt} = -g^{\mu{j}}\Dp{h}{x^{{j}}}
 \end{equation}
 which start from a critical point $b$ and end at a critical point $a$
(with $h(a)$ necessarily less than $h(b)$) are discrete (and
finite in number).

In \cite{Witten82} Witten first shows that, if $H_{u}$ is the Hamiltonian
obtained when $h$ is scaled to $uh$, then corresponding to each critical
point $a$ of $h$ of degree $p$ there is exactly one eigenstate
$\psi_a(x,\eta)$ of $H_{u}$ with the following properties:
it has low eigenvalue, in the sense that it does not tend to infinity
with $u$, it is a form of degree $p$ (that is, it is of degree $p$ in
$\eta$) and it is concentrated near $p$.
Furthermore no there are no other low-eigenvalue eigenstates of $H_{u}$.
The simplest way to see this is to observe that in Morse coordinates and
using the free choice of metric to pick a metric which is Euclidean on
$N_a$ the Hamiltonian
$H_{u}$ splits into a bosonic part  which is simply an oscillator Hamiltonian
and a fermionic part which reflects the index $p$ and is easily seen to
have
$p$-form eigenfunctions. A fully rigorous proof of the result is given
by Simon et al in \cite{Simetal}.

Now $H_{u}$ and $d_{u} \equiv e^{-hu} d e^{hu}$ commute, and so if $\psi$
is a low eigenvalue eigenstate of $H_{u}$, then $d_{u}\psi$ is also a low
eigenvalue eigenstate.  Hence for each critical point $a$ of index $p$
 \begin{equation}
  d_{u} \psi_{a} = \Sum{b \Mbox{\ critical of index\ } p+1}{}
    C_{ab} \psi_{b}
 \end{equation}
for some real numbers $C_{ab}$ (which will in due course be determined).
We thus see that the real cohomology of $M$ can be modelled by
$p$-cochains
$\sigma$ taking the form
 \begin{equation}\label{COCHAINeq}
  \sigma = \sum_{a \Mbox{\ critical}} \sigma_{a} \psi_{a}
 \end{equation}
where the coefficients $\sigma_{a}$ are real numbers, and the coboundary
operator is $d_{u}$.

The coefficients $C_{ab}$ which determine the action of $d_{u}$
may be determined from the matrix elements $d_{u\,2} \exp -H_u
t(A,B)$ in the large $u$ limit. (The subscript $_2$ indicates that
the derivative is taken with respect to the second argument.) We
may take an orthonormal basis of eigenstates of $H_{u}$ consisting
of the low eigenvalue states $\psi_{u\,a}$ with eigenvalues
$\lambda_{u\,a}$ for each critical point $a$ and further
eigenstates $\psi_{u\,n}, n =1, \dots$ with eigenvalues
$\lambda_{u\,n}$ all of which tend to infinity with $u$ and, for
fixed $u$, tend to infinity with $n$. Then we can express the
kernel of the evolution operator as
 \begin{equation}\label{EEXPeq}
  \exp -H_{u}t(X,Y) = \sum_{a \Msbox{critical}}
  e^{-\lambda_{u\,a}t} \Cj{\psi}_{u\,a}(X) \psi_{u\,a}(Y)
 +\sum_{n} e^{-\lambda_{n}t} \Cj{\psi}_{n}(X) \psi_{n}(Y).
 \end{equation}
For large $u$ at leading order this becomes
 \begin{equation}\label{EEXPLeq}
  \exp -H_{u}t(X,Y) = \sum_{a \Msbox{critical}}
   \Cj{\psi}_{u\,a}(X) \psi_{u\,a}(Y)
  \end{equation}
so that at leading order for large $u$
 \begin{equation}\label{CVALeq}
  d_{u\,2} \exp -H_{u}t (A,B) = C_{ab} \Cj{\psi_a}(A){\psi_b}(B)
 \end{equation}
where for notational simplicity we have omitted the subscripts $u$.

To evaluate this expression we use the Feynman-Kac-It\^o formula
(\ref{FKIeq}), together with the explicit form (\ref{MORSECOORDeq}) of
$h$ in the neighbourhood of a critical point of $h$.  We choose a metric
which globally satisfies the Smale transversality condition (so that
classical curves $\Gamma_{ab}$ of steepest descent from the critical
point $a$ to the critical point $b$ are finite in number and discrete)
and also one which is Euclidean in a set of Morse coordinates around each
critical p 

Before proceeding further it is useful to introduce some specific
coordinate systems. For each critical point $c$ in $M$ we will
choose on $N_c$ a fiducial set of Morse coordinates
$x_{[c]}^{\mu}$ and fermionic partners $\eta_{[c]}^{\mu}$.
Additionally for each steepest descent curve
 $\Gamma_{ab}$ joining the pair of
critical points $a$ and $b$ with indices $p$ and $p+1$ respectively we
will choose a coordinate neighbourhood $U\Gab$ which contains
 $N_a\cup N_b \cup U\Gab$  with coordinates
$x\Gab,\eta\Gab$  such that $\Gamma_{ab}$ lies along $x\Gab^{n-p}$
while $x\Gab,\eta\Gab$ match $x\Gb,\eta\Gb$ on $N_b$ apart from
possible rotations, and also match  $x\Ga$ on $N_a$ apart from
possible rotations and  (necessarily) a translation in the
$x^{n-p}$ coordinate with $x^{n-p}\Gab=x\Ga^{n-p} + k_a$ for some
positive constant $k_a$. Eventually, when the coordinate systems
are reconciled,  sign factors will be produced.

Within $N_a$ the Hamiltonian then has the form
\begin{eqnarray}
  \Hamu &=& \sum_{{i}=1}^n \Bigg[
   \Frac12 \Big(-\Frac{\partial^2}
   {\partial x\Gab^{{i}} \partial x\Gab^{{i}}}    +
   u^2 (x\Gab^{{i}}-a\Gab^{{i}})(x\Gab^{{i}}-a\Gab^{{i}})\Big)
    \End
 &&\qquad \quad + \quad \Frac{i}{2}u \sigma_{i}
    (\eta\Gab^{{i}}\pi\Gab{}_{{i}}-\pi\Gab{}_{{i}}\eta\Gab^{{i}}) \Bigg].
\end{eqnarray}

 The bosonic and fermionic parts commute so that their heat
kernels may be considered separately; the bosonic part is the
Harmonic oscillator Hamiltonian whose heat kernel is given by
Mehler's formula \cite{Simon}, while the fermionic part is (apart
from sign factors $\sigma_{i}$) the fermionic oscillator whose
heat kernel is given in \cite{GBM}. If $x$ is near $a$ and in
$N_a$ then at leading order for large $u$
 \begin{eqnarray}
  \lefteqn{ \exp -\Hamu{}t (A,X) =_{{\rm def}} \Meh(A,X) } \End
  &=&      \left( \frac{u}{\pi}\right)^{n/2}
     \Exp{- \Frac12 u (x\Gab{}-k_a)^2}
     \prod_{{i}=1}^{n-p}( - \alpha\Gab^{{i}})\prod_{{j}=n-p+1}^{n}
      \eta\Gab^{{j}}  \End
 \end{eqnarray}
where $X$ is a point over $x$ in $N_{a}$, with  coordinates
$(x\Gab,\eta\Gab)$.

Next we calculate $\exp -\Hamu t(A,X)$ for $x$ near $\Gamma_{ab}$ using
the Feynman-Kac-It\^o formula (\ref{FKIeq}). In this case the steepest
descent  curve (satisfying (\ref{STeq})) from $x$ approaches $a$ very
fast. Thus after very small time $\delta t$ the path $\tilde{x}_{\delta
t}$ is almost certainly near $a$, so that to leading order in $u$ we have
a contribution from
$\Gamma_{ab}$ of
  \begin{eqnarray}\label{GABKEReq}
  \lefteqn{ \exp -\Hamu t(A,X)\Gab{}=\exp -\Hamu \delta t
  \exp -\Hamu (t-\delta t) (A,X) }  \End
  &=&  \Exs  \Bigg[ \Exp{- u( h(x)-h(\xm_{\delta t}))} \End
\times && \exp\Bigg( \Intdt  u\Dctwoh{\xm_s}{{i}}{{{j}}}
  i g^{{k}{j}}(\xm_s)  \Fmpos^{{i}}_s \Fmmom_{s\,{k}}
   \End
&& {}+ \Curv{}{}{{i}}{{j}}(\xm_s) \Fmpos^{{i}}_s \Fmmom_{{j}\, s}
 +\frac12 R_{{i}{k}}^{{k}{j}}(\xm_s)
 \Fmpos^{{i}}_s \Fmpos^{{k}}_s \Fmmom_{{k}\, s}\Fmmom_{{j}\, s}
          \,ds \Bigg)
 \Meh(a\Gab,\alpha\Gab,\xm_{\delta t},\Fmpos_{\delta t}) \Bigg]  \End
  &=& \left( \frac{u}{\pi}\right)^{n/2}\Exp{ - u(h(x)-h(a))}
   \prod_{{i}=1}^{n-p}( - \alpha\Gab^{{i}})\prod_{{j}=n-p+1}^{n}
      \eta\Gab^{{j}}. \End
 \end{eqnarray}
Here we have used the fact that the operator $\eta^{i}\pi_{i}$
which corresponds to the term $g^{{k}{j}}(\xm_s)  \Fmpos^{{i}}_s
\Fmmom_{s\,{k}} $ in the path integral has zero eigenvalue on the kernel
 $\exp-\Hamu t (A,\xm_{\delta t},\Fmpos_{\delta t})$ when $x$ lies on $\Gamma_{ab}$.

To calculate $\Dut \exp -\Hamu t(A,B)$ we cannot take the derivative of
the separate contributions from each $\Gamma_{ab}$ using (\ref{GABKEReq})
because as we vary $x$ around $b$ we will jump from one $\Gamma_{ab}$ to
another. To avoid this difficulty we note that
  \begin{eqnarray}
  \lefteqn{\Dut \exp -\Hamu t (A,B) }\End
  &=& \Dut\exp-\Hamu s \,  \exp -\Hamu (t-s)(A,B)  \End
   &=& \int_{M} d^n x d^n \eta   \exp-\Hamu (t-s) (A,X) \Dut \exp -\Hamu s(X,B).
  \label{FACeq}\end{eqnarray}
Because of the concentration of $\Dut\exp -Hs(X,B)$ near $b$ we can
integrate over $\Real^n$ rather than $M$ using the form of
$\exp -Hs(X,B)$ which is approximately true for large $u$ on
$N_b$; although ultimately we will obtain a result independent of
$s$ and $t$, at this stage we must use Mehler's formula in full
(including terms of order $e^{-us}$ whose equivalent we could neglect
near $a$ for our purposes) because it is not the zero mode of $\Hamu$
which will contribute to $d\psi_a(b)$ at leading order. Thus for $x$ and
$y$ near $b$ we use
 \begin{eqnarray}
 \lefteqn{\exp -\Hamu s(X,Y) = \left( \frac{u}{\pi}\right)^{n/2}
 \Exp{-  \Frac12 u \left(x\Gab^2 \Csh \right) +u \,\frac{x\Gab y\Gab}{\sinh us}} }\End
 &&\qquad\qquad\times \,\prod_{{i}=1}^{n-p-1}
  \left( \phi\Gab^{{i}} e^{- us} - \eta\Gab^{{i}}  \right)
 \prod_{{j}=n-p}^{n} \left( \phi\Gab^{{j}}  - \eta\Gab^{{j}}e^{- us}
 \right).
 \End
 \end{eqnarray}
where and $(x\Gab,\eta\Gab)$, $(y\Gab,\phi\Gab)$ are the
coordinates of $X$ and $Y$ respectively, so that  the term of
$\Dut\exp -\Hamu s(X,B) $ of leading order in $u$ is
 \begin{equation}
 \left( \frac{u}{\pi} \right)^{n/2}
 u\,\frac{x^{n-p}}{\sinh us} \Exp{-  \Frac12 u \left(x\Gab^2 \Csh \right)}
  \quad\,\prod_{{i}=1}^{n-p} \eta\Gab^{{i}}
 e^{-us} \prod_{{j}=n-p}^{n}  \beta\Gab^{{j}}  .
  \end{equation}
Using (\ref{FACeq}) gives
 \begin{eqnarray}
\lefteqn{\Dut  \exp -\Hamu t (A,B)}\End
 &=&\int_{\Real^n} d^{n}x\Gab \,
 \left(\frac{u}{\pi}\right)^{n}\theta(x\Gab^{n-p})
 u\, \frac{e^{-us}}{\sinh us} x\Gab^{n-p}
 \prod_{{i}=1}^{n-p} \left(  - \alpha\Gab^{{i}} \right)
 \prod_{{j}=n-p}^{n}  \beta\Gab^{{j}}\End
 &&\times\Exp{-\Frac12 u x\Gab^2 \Csh}\exp-u(h(x)-h(a)) \End
 &=&
 \int_0^{\infty} dx\Gab^{n-p}
 \left(\frac{u}{\pi}\right)^{(n+1)/2}u\, \frac{e^{-us}}{\sinh us} x\Gab^{n-p}
 \prod_{{i}=1}^{n-p} \left(  - \alpha\Gab^{{i}} \right)
 \prod_{{j}=n-p}^{n}  \beta\Gab^{{j}}\End
  &&\times \Exp{-\Frac12 u x\Gab^2 \left(\Csh-1 \right)}\exp-u(h(b)-h(a)) \End
 &=&\left(\frac{u}{\pi}\right)^{(n+1)/2}
 \prod_{{i}=1}^{n-p} \left(  - \alpha\Gab^{{i}} \right)
 \prod_{{j}=n-p}^{n}  \beta\Gab^{{j}}
 \exp-u(h(b)-h(a))
 \end{eqnarray}
at leading order in $u$. Here the $\theta$-function occurs because
the contribution from $\Gamma_{ab}$ to $\exp -\Hamu (t-s)$ is zero
on the side of $b$ away from $a$. Now using equation
(\ref{CVALeq}) and the fact that
 \begin{equation}
 \Hstar\psi_{a}(A)= \left( \frac{u}{\pi}\right)^{n/4}
 \prod_{{i}=1}^{n-p}\alpha\Ga^{{j}},
  \quad
 \psi_{b}(B)= \left( \frac{u}{\pi}\right)^{n/4}
 \prod_{{j}=n-p}^{n}\beta_{[b]}^{{j}}
  \end{equation}
we see that
 \begin{equation}
 c_{ab} = \left( \frac{u}{\pi}\right)^{1/2}
   \exp-u(h(b)-h(a))\sum_{\Gamma_{ab}} (-1)^{\sigma\Gab}
 \end{equation}
where $(-1)^{\sigma\Gab}$ is a sign factor which comes from
transforming between the $[a]$ and $[b]$ coordinates and the
$[\Gamma_{ab}]$ coordinates.

If (again following Witten \cite{Witten82}) we rescale each
$\psi_c$ to $\psit_c=e^{-uh(c)}\psi$, and also define
$\tilde{d_{u}}= \sqrt{\frac{\pi}{u}}d_u$, we obtain
 \begin{equation}
  \tilde{d}_u\psit_a
  = \sum_{\Gamma_{ab}} (-1)^{\sigma_{\Gamma_{ab}}} \psit_{b}
  \end{equation}
which coincides with the geometrical approach using ascending and
descending  spheres.
%
%

 \end{document}